\documentclass[prd,aps,twocolumn,superscriptaddress,preprintnumbers,nofootinbib,showpacs]{revtex4}
\usepackage{graphicx}
\usepackage{exscale}
\usepackage[intlimits]{amsmath}
\usepackage{amsfonts}
\usepackage{amssymb,amscd}
\usepackage{epsfig}
\usepackage{pstricks}

\newcounter{commentdepth}
\newcommand{\be}{\begin{equation}}
\newcommand{\ee}{\end{equation}}

\newcommand{\1}{\ensuremath 1\!\!1}
\renewcommand{\d}{\ensuremath\mathrm{d}}

\newcommand{\E}{\ensuremath\mathrm{e}}
\newcommand{\eps}{\ensuremath\varepsilon}

\newcommand{\I}{\ensuremath \mathrm{i}}

\newcommand{\norm}[1]{\ensuremath\left| #1 \right|}

\newcommand{\V}[1]{\boldsymbol{#1}}

\newcommand{\beqa}{\begin{eqnarray}}
\newcommand{\eeqa}{\end{eqnarray}}
\newcommand{\beq}{\begin{equation}}
\newcommand{\eeq}{\end{equation}}

\newcommand{\p}{\partial}

\newcommand{\reftitle}[1]{}

\begin{document}

\preprint{\parbox[t]{\textwidth}
{\small DATE [\today] \hspace*{5mm} IDENTIFIER? \hfill ArXiv:0911.5435 [hep-ph]}}

\title{Infrared Critical Exponents in Finite-Temperature Coulomb Gauge QCD}

\author{Klaus~Lichtenegger}
\affiliation{Institut f\"ur Physik, Karl-Franzens-Universit\"at Graz, 8010 Graz, Austria}

\author{Daniel~Zwanziger}
\affiliation{Physics Department, New York University, New York, NY 10003, USA}

\begin{abstract}
We investigate the infrared critical exponents of Coulomb
gauge Yang-Mills theory in the limit of very high temperature.
This allows us to focus on one scale (the spatial momentum) since
all but the lowest Matsubara frequency decouple from the deep infrared.

From the first-order Dyson-Schwinger equations in a bare-vertex
truncation we obtain infrared exponents which correspond to
confining or overconfining (yet mathematically well-defined)
solutions. For three spatial dimensions the exponents are close
to what is expected for a linearly rising color-Coulomb potential.
\end{abstract}


\maketitle

\newlength{\xskip}
\setlength{\xskip}{-1.5cm}
\newlength{\xskipp}
\setlength{\xskipp}{-1.32cm}

\section{Introduction}
\label{sec:IRCoulomb_intro}

Crucial features of QCD are believed to be encoded in the infrared
behaviour of its Greens functions. Indeed, the greatest unsolved
problems of the theory -- confinement, dynamical chiral symmetry
breaking and the emergence of a mass gap -- presumably have
their origin in the infrared, where interactions are strong, the
coupling is large and perturbation theory breaks down.

Thus it is extremely valuable to have methods at hand which allow
us to study this sector. As argued in~\cite{von Smekal:1997is,
von Smekal:1997vx}, in the deep infrared, far away from all other
scales ($\Lambda_{\text{QCD}}$ or quark masses), the theory
should be conformal and propagators $D$ should thus exhibit
an asymptotic power-law behaviour,
\beq
  D_i(p) \sim c_i(\mu)\,(p^2)^{\delta_i}
\eeq
with infrared critical exponents $\delta_i$ (and coefficients
$c_i$ which contain some power of a renormalization scale $\mu$.)
These exponents can be extracted from the Dyson-Schwinger
equations (DSEs), the equations of motion for a quantum field
theory.

This endeavor has been pursued with great success
in the Landau gauge~\cite{Alkofer:2000wg}, where, in accordance
with Gribov's confinement scenario~\cite{Gribov:1977wm, Zwanziger:1989mf,
Zwanziger:1993}, the infrared suppression of the gluon propagator
could be traced back to the enhanced divergence of the ghost
propagator.\footnote{Note that some recent lattice studies question
these results. While those concerns are certainly to be taken
seriously, it would seem overly hasty to dismiss the results obtained
by functional methods. In particular, since the finite size of a
lattice corresponds to an infrared cutoff in momentum space,
huge lattices are required to perform a reliable extrapolation to
infinite volume, and there is still an ongoing debate about systematic
errors~\cite{Cucchieri:2007md, vonSmekal:2007ns}. In general,
there is a discussion about the role of the ``massive'' as opposed
to the ``scaling'' solution~\cite{Fischer:2008yv}.}

Unfortunately, other gauges seem to be more difficult to access.
In the Coulomb gauge, which is of particular interest for the
present authors, the SO(3,1) symmetry of spacetime is reduced
to SO(3), thus a propagator $D(p)$ generically depends on
two physical momentum scales $p_0^2$ and $\V{p}^2$.

Accordingly various different infrared limits have to be distinguished:
Even if one has $\V{p}^2\to 0$ and $p_0^2\to 0$, one could have
completely different behaviour depending on the ratio
$\frac{p_0^2}{\V{p}^2}$.\footnote{In general one encounters
all problems present for functions of two real variables,
where not only in general
$\lim_{x\to 0}\lim_{y\to 0}f(x,y)\ne\lim_{y\to 0}\lim_{x\to 0}f(x,y)$,
but the limit $\lim_{(x_n,y_n)\to(0,0)}f(x_n,y_n)$
may even depend on the precise path on which the origin is approached.}
Even in a quasi-instantaneous approximation (where all diagrams
without at least one instantaneous propagator have been neglected)
infrared critical exponents have turned out to be rather elusive
objects~\cite{Alkofer:2009dm}.

Thus in this article, instead of working with the ground-state
theory, we study QCD (or rather SU($N$) gauge theory) in
the limit of very large temperature, where certain simplifications
occur. At first glance one may wonder how useful this
could be, taking into account that according to common lore,
at large temperatures QCD becomes ``deconfined'', chiral
symmetry is restored and the mass gap is gone.

However, as discussed in section~1 of~\cite{Lichtenegger:2008mh}
(where additional references are given), there is increasing evidence that this
picture is likely to be incomplete or even wrong. Perturbative~\cite{Linde:1980ts,
Gross:1980br}, lattice~\cite{Boyd:1996bx}, functional~\cite{Maas:2004se,
Maas:2005hs} and experimental~\cite{Shuryak:2004cy}
results suggest that the infrared sector and bound states
play an essential role also at very high temperature, so
that there is no ``deconfined phase'' in the strict sense.

Thus one can hope to obtain useful information
about the infrared sector even in the case of very
large temperatures, which we will study in the following.

\subsection*{Organization of this Article}

The article is organized as follows:
In sec.~\ref{sec:IRExpFinT_LocalAction} we give the action of
Coulomb-gauge Yang-Mills theory for finite (and in particular
extremely high) temperature; in sec.~\ref{sec:IRExpFinT_DefProp}
we define propagators and proper two-point functions and
discuss the relation between them.

In sec.~\ref{sec:IRExpFinT_TruncDSEs} we state the
Dyson-Schwinger equations of the theory, which are
examined more closely in a bare-vertex truncation.
Since these equations contain infrared-divergent
integrals, we discuss the topic of these singularities
in sec.~\ref{sec:IRExpFinT_IRDiv}.

We define the infrared critical exponents in
sec.~\ref{sec:IRExpFinT_IRDef} and employ power-law
ans\"atze in sec.~\ref{sec:IRExpFinT_IRDSEs} in
order to obtain the infrared asymptotic Dyson-Schwinger
equations.

The exponents are constrained analytically and
determined numerically in sec.~\ref{sec:IRExpFinT_DetCritExp}.
(The evaluation of power-law integrals is discussed
in appendix~\ref{app:IRExpFinT_PowerLaw}.)
In addition we can also determine relations
between the propagator coefficients, which
are derived in sec.~\ref{sec:IRExpFinT_RelbCoeff}.

In sec.~\ref{sec:IRExpFinT_RangVal} we try to estimate the
range of validity of the approach followed in this article, and
in sec.~\ref{sec:IRExpFinT_Summary} we discuss the results
obtained for the infrared critical exponents and give a brief summary.

\section{Local action}
\label{sec:IRExpFinT_LocalAction}

Our starting point is the Yang-Mills action in $d=s+1$ dimensions,
gauge-fixed to the Coulomb gauge, which is particularly well-suited
for studies of finite-temperature field theory,
\beq
  S_{\text{YM,Coul}} = \int \d^{s+1}x\left( \frac14
    F_{\mu\nu}^2 - \I(\p_ib)A_i + (\p_i\bar{c})D_ic\right)
  \label{eq:IRExp_YMCoulomb}
\eeq
where $F_{\mu \nu} = \p_\mu A_\nu - \p_\nu A_\mu + g A_\mu \times A_\nu$,
and $(A_\mu \times A_\nu)^a \equiv f^{abc} A_\mu^b A_\nu^c$;
the gauge-covariant derivative is given by $D_i c = D_i[A] c \equiv \p_i c + g A_i \times c$.
We will now modify the action~\eqref{eq:IRExp_YMCoulomb} in four ways:
\begin{itemize}
\item We apply the on-shell formalism, so the Nakanishi-Lautrup field
  $b$ is integrated out in order to directly impose the transversility
  condition
  \beq
    \p_i A_i = 0\,.
  \eeq
\item We turn to finite temperature, so the temporal integral has the limits $\int_0^\beta dx_0$, where $\beta = {1 \over T}$ and $T$ is the temperature.  Integrals over $k_0$ will be
  replaced by a sum over Matsubara frequencies, $\int dk_0 \to T \sum_n$. 
  
  \item  We neglect all but the 0th Matsubara frequency,\footnote{Note that
the Linde problem~\cite{Linde:1980ts} has its origin in the zeroth
Matsubara frequency as well.} which is the same as dropping all time derivatives, $\partial_0 \rightarrow 0$, in the action, and replacing $\int_0^\beta dx_0 \to {1 \over T}$, so the action simplifies to
\beq
\label{secondorder}
S_1 = {1 \over T} \int \d^s x\left( {1 \over 2} (D_iA_0)^2 + \frac14
    F_{ij}^2 + \p_i\bar{c}D_ic\right).
\eeq  
  
  \item We rewrite the theory in the \emph{first-order formalism}, 
    by  introducing a new field $\pi_i^a$ by a Gaussian identity, so the action reads 
  \begin{align}
  S_2 &= {1 \over T} \int \d^sx \Big[ \I \pi_i(-D_iA_0)
  + \frac12 \pi_i^2  + \frac14F_{ij}^2
  + \partial_i\bar{c}D_ic \Big]\,.
\end{align}
The new field can be interpreted as the
momentum  conjugate to $A_i^a$ and thus plays the role of a
  color-electric field.  It can be decomposed into
  transverse and longitudinal parts
  \beq
    \pi_i = \pi_i' - \p_i \varphi,
  \eeq  
  where $\pi_i'$ is transverse, $\p_i \pi_i' = 0$, which gives the action that will be used to derive the DSEs,
\begin{align}
  S &= {1 \over T} \int \d^sx \Big[ \I (\pi_i'-\partial_i\varphi)(-D_i A_0)
  + \frac12 \left(\pi_i'\right)^2 \nonumber \\
  & \qquad + \frac12 (\partial_i\varphi)^2 + \frac14F_{ij}^2
  + \partial_i\bar{c}D_i c \Big]\,.
  \label{eq:IRfinT_firstaction}
\end{align}
\end{itemize}

\section{Definition of propagators and proper 2-point functions}
\label{sec:IRExpFinT_DefProp}

If we confine ourselves to the zero Matsubara frequency,
propagators only depend on the spatial momentum. The
propagators of the transverse fields are defined as
\begin{align}
\langle A_i^a (x) A_j^b(y) \rangle &= \int \frac{d^s k}{(2\pi)^s} \ 
  \E^{\I k\cdot (x-y)} \delta^{ab} P^T_{ij}(k) D_{AA}(k)\,, \nonumber \\
\langle {\pi'}_i^a (x) {\pi'}_j^b(y) \rangle &= \int \frac{d^s k}{(2\pi)^s} \
  \E^{\I k\cdot (x-y)} \delta^{ab} P^T_{ij}(k) D_{\pi \pi}(k)\,,
\end{align}
where $P^T_{ij}(k)$ is the transverse projector,
\begin{equation}
  P^T_{ij}(k) = \delta_{ij} - \frac{k_i k_j}{k^2}\,.
\end{equation}

The propagator $\langle \pi'_i (x) A_j(y) \rangle$ is proportional to $k_0$
both at tree-level and for the power-law ans\"atze employed in
sec.~\ref{sec:IRExpFinT_IRDef}. Thus it has vanishing zero-Matsubara
component in this context, and will be neglected in the asymptotic infrared limit,
\begin{equation}
  \langle \pi'_i (x) A_j(y) \rangle = 0\,.
  \label{eq:IRExp_mixedvanish}
\end{equation}
This removes the mixing of the transverse fields, so the proper functions
are given as the one-dimensional inverse of the propagators,
\begin{align}
  \Gamma_{\bf AA}(k) &= \frac{D_{\pi'\pi'}}{D_{\bf AA}D_{\pi'\pi'}-D_{\mathbf{A}\pi'}^2}
     \to \frac1{D_{\bf AA}(k)}\,, \\
  \Gamma_{\pi'\pi'}(k) &= \frac{D_{\bf AA}}{D_{\bf AA}D_{\pi'\pi'}-D_{\mathbf{A}\pi'}^2}
     \to \frac1{D_{\pi'\pi'}(k)}\,.
\end{align}
On the other hand the scalar Bose fields do mix.
Their propagators are defined by
\begin{eqnarray}
\langle A_0^a (x) A_0^b(y) \rangle & = &
   \int \frac{d^s k}{(2\pi)^s} \ \E^{\I k\cdot (x-y)} \delta^{ab} D_{A_0 A_0}(k)\,,  
\nonumber   \\
\langle A_0^a (x) \varphi^b(y) \rangle  & = &
  \int \frac{d^s k}{(2\pi)^s} \ \E^{\I k\cdot (x-y)} \delta^{ab} D_{A_0 \varphi}(k)\,,
\nonumber   \\
\langle \varphi^a(x) \varphi^b(y) \rangle  & = &
  \int \frac{d^s k}{(2\pi)^s} \ \E^{\I k\cdot (x-y)} \delta^{ab} D_{\varphi \varphi}(k)\,,
\end{eqnarray}
the Faddeev-Popov ghost propagator is defined by
\begin{equation}
  \langle  c^a(x) \bar c^b(y) \rangle
     = \int \frac{d^s k}{(2\pi)^s} \ \E^{\I k\cdot (x-y)} \delta^{ab} D_{c \bar c}(k).
\end{equation}
While the inversion of the ghost propagator (in order to obtain the proper $2$-point function)
is simple,
\beq
  \Gamma_{\bar c  c}(k) = { 1 \over D_{c \bar c(k)} }\,,
\eeq
for the other scalar fields the proper 2-point functions are
two-dimensional matrix inverses of the propagators,
\begin{align}
  \Gamma_{A_0 A_0}(k) = & { D_{\varphi \varphi}(k) \over \Delta (k)}\,,
    \hspace{.5cm}  \Gamma_{\varphi \varphi}(k) = { D_{A_0 A_0}(k) \over \Delta (k)}\,, \nonumber  \\
  &\Gamma_{A_0\varphi}(k) = -{D_{\varphi A_0}(k) \over \Delta (k)},
\end{align}
where 
\begin{equation}
\Delta \equiv D_{\varphi \varphi} D_{A_0 A_0} - D_{A_0 \varphi}^2.
\end{equation}

\section{Truncated Dyson-Schwinger equations}
\label{sec:IRExpFinT_TruncDSEs}

The derivation of the Dyson-Schwinger equations (DSEs) for the theory
described by~\eqref{eq:IRfinT_firstaction} is straightforward, yet tedious.
(Details of the derivation will be discussed in~\cite{KlausPhD}.)
To simplify this endeavour we neglect the cubic and quartic pieces of
$F_{ij}$ in~\eqref{eq:IRfinT_firstaction}, since the scalar fields $A_0$,
$\varphi$, $c$ and $\bar{c}$ are expected to be dominant in the infrared;
loops containing a three- or four-gluon vertex (and the the corresponding
amount of transverse propagators) are supposed to be subdominant..

\begin{figure*}
\begin{center}
  \includegraphics[width=15.8cm]{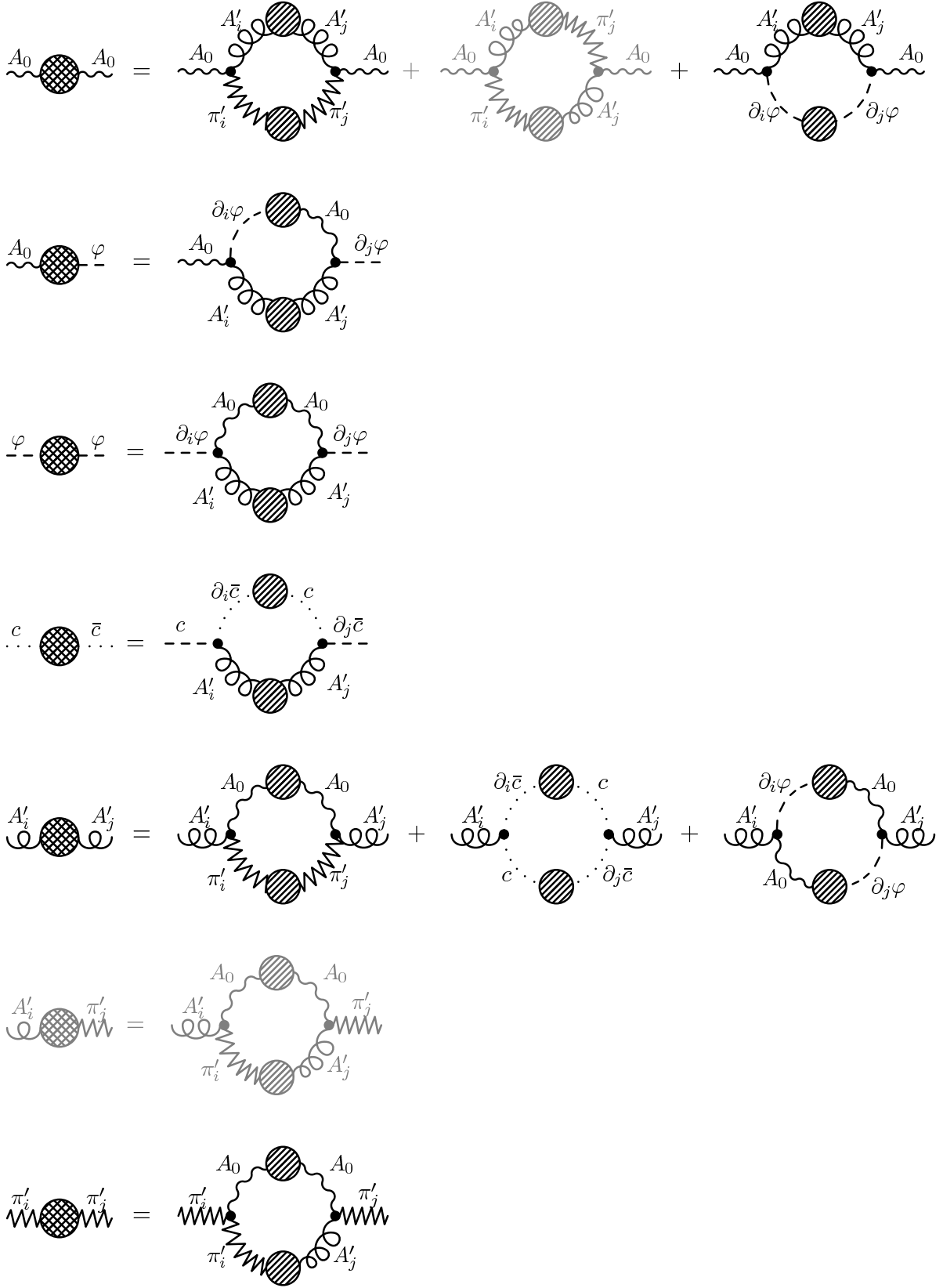}
\end{center}
\caption{The system of bare vertex truncated Dyson-Schwinger equations
in the first-order formalism, where the tree-level terms have (for the transverse
fields) been neglected or (for the scalar fields) removed by imposing the horizon
condition. Prefactors and signs have been absorbed in the graphs. Diagrams
and equations drawn in gray drop out in the
approximation~\eqref{eq:IRExp_mixedvanish}.}
\label{fig:IRCoulomb_firstDSEs}
\end{figure*}

To further simplify the equations (and since little is known about the
dressed vertices of this theory anyway), we employ a truncation
in which dressed vertices are replaced by bare ones. (Note that
in~\cite{Alkofer:2009dm} a more general ansatz for the vertices did
not change the general picture.) A graphical representation of the
resulting DSEs is given in Figure~\ref{fig:IRCoulomb_firstDSEs}.

With these truncations, the equation for  $\Gamma_{\bf AA}$ reads
\beq
  P_{ij}^T(k)\Gamma_{\bf AA}(k) = P_{ij}^T(k) \ k^2 + g^2 TN P_{im}^T(k) \ I_{mn} \ P_{nj}^T(k).
  \label{eq:DSEGammaATAT}
\eeq
where the temperature $T$ stems from the Matsubara sum
(of which we keep only the zeroth term). The loop integral $I_{mn}(k)$
is sandwiched between transverse projectors $P^T(k)$, and is given by
\beqa
  I_{mn}(k) && \equiv  \int {d^s p \over (2\pi)^s} 
  \Big[ D_{\pi_m \pi_n}(p) \ D_{A_0 A_0}(p+k) + p_m p_n
\nonumber   \\   \nonumber   
&& \times   \Big( D_{\varphi A_0}(p) D_{A_0 \varphi}(p+k)
+ D_{c \bar c}(p) \ D_{c \bar c}(p+k) \Big) \Big], 
\\ && \ 
\eeqa
where the propagator of the color-electric field has the decomposition
\beq
  \label{decompDpipi}
  D_{\pi_m \pi_n}(p) = P^T_{mn}(p)  D_{\pi' \pi'}(p) +  p_m p_n D_{\varphi \varphi}(p).
\eeq
The DS equation for $\Gamma_{\pi' \pi'}$ reads
\beq
  P_{ij}^T(k)\Gamma_{\pi' \pi'}(k) = P_{ij}^T(k) + g^2 TN P_{im}^T(k) J_{mn} P_{nj}^T(k)
\eeq
where the loop integral is given by
\beq
  J_{mn}(k) \equiv \int {d^s p \over (2\pi)^s} \ P^T_{mn}(p) D_{\bf AA}(p) \ D_{A_0 A_0}(p+k).
\eeq

We further obtain the equation for $\Gamma_{A_0 A_0}$,
\beq
  \Gamma_{A_0 A_0}(k) =  g^2 TN  \int {d^s p \over (2\pi)^s} \  P^T_{ij}(p)  D_{\bf AA}(p) D_{\pi_j \pi_i}(p+k),
  \label{eq:DSEGamma00}
\eeq
where $D_{\pi_j \pi_i}$ is given in (\ref{decompDpipi}).
The DS equation for $\Gamma_{\varphi \varphi}$ reads
\begin{align}
\Gamma_{\varphi \varphi}(k) &= k^2 + g^2 TN  \int {d^s p \over (2\pi)^s} \ k_i k_j P^T_{ij}(p) D_{\bf AA}(p) 
\nonumber  \\
&\hspace{2.5cm} \times D_{A_0 A_0}(p+k)\,,
\end{align}
the DS equation for $\Gamma_{\varphi A_0}$ is given by
\beqa
\Gamma_{\varphi A_0}(k) = i k^2 + g^2 T N 
\int {d^s p \over (2\pi)^s} \
P^T_{ij}(p) D_{\bf AA}(p) 
\nonumber  \\
\times k_i k_j  D_{A_0 \varphi}(p+k).
\eeqa
Finally the equation for $\Gamma_{\bar c c}$ reads
\beqa
\Gamma_{\bar c c}(k) = k^2 - g^2 T N 
\int {d^s p \over (2\pi)^s} \
P^T_{ij}(p) D_{\bf AA}(p)
\nonumber  \\ 
\times k_i k_j  D_{c \bar c}(p+k).
\label{eq:DSEGammacc}
\eeqa

The tree-level terms in equations~\eqref{eq:DSEGammaATAT}
to~\eqref{eq:DSEGammacc} do not directly affect the infrared
asymptotic behaviour for esentially two reasons:

\begin{itemize}
\item The scalar fields are expected to be infrared-enhanced.
  Accordingly we impose the horizon condition~\cite{Zwanziger:1989mf,
  Zwanziger:1993} on the Faddeev-Popov ghosts and (since we expect
  at least qualitatively analogous behaviour) also on the bosonic fields.
  As a consequence, the tree-level part in~\eqref{eq:DSEGamma00}
  to~\eqref{eq:DSEGammacc} is cancelled by quantum fluctuations.

\item The DS equations for the transverse fields contain at least
  one (uncancelled) loop with at least one scalar propagator, which
  will -- due to infrared enhancement -- dominate the tree-level part. 
\end{itemize}

\section{Infrared Divergence in DSE}
\label{sec:IRExpFinT_IRDiv}

In Coulomb gauge, the time-time component of the gluon propagator has an instantaneous part,
\beq
  \label{vcoul}
  - g^2 D_{A_0A_0}(x,t) = V_{\rm coul}(r) \delta(t) + V_{\rm non-inst.}(x,t),
\eeq
known as the color-Coulomb potential $V_{\rm coul}(r)$, where $r = |x|$.  When the gauge-invariant potential introduced by Wilson is confining, $\lim_{r \to \infty}V_{\rm W}(r) = \infty$, it provides a lower bound on the color-Coulomb potential asymptotically at large $r$, $V_{\rm coul}(r) \geq V_{\rm W}(r)$, summarized by ``no confinement without Coulomb confinement" \cite{Zwanziger:2003}.  In the confining phase we expect the Wilson potential to rise linearly, $V_{\rm W}(r) \sim \sigma r$, where $\sigma$ is the physical string tension, and in this case the color-Coulomb potential rises (at least) linearly, $V_{\rm coul}(r) \sim \sigma_{\rm coul} r$, where $\sigma_{\rm coul}$ is the Coulomb string tension and $\sigma_{\rm coul} \geq \sigma$.  Moreover in Coulomb gauge $gA_0$ is a renormalization-group invariant~\cite{Zwanziger:2003}, which implies that $V_{\rm coul}(r)$ is also a renormalization-group invariant.  Thus, as long as $\sigma_{\rm  coul}$ is finite, it has a well-defined physical value in MeV which has been calculated in lattice gauge theory~\cite{Greensite:2004ke, Nakagawa:2006fk,Nakagawa:2007zzc, Nakagawa:2007zzb,Voigt:2008rr}.

	The linearly rising potential $r$ corresponds in momentum space to $1/k^4$, so one encounters infrared divergences of the type $\int d^3k/k^4$ in the DS equations.  To address this problem, we observe that the DS equations are originally derived in position space [from the functional identity for  $ \delta \Gamma / \delta A(x)$], and moreover these equations remain free of infrared divergences when written in position space, even in the presence of long-range potentials.  Indeed a loop integral such as 
\beq
\label{convolution}
L(p) = \int {d^sk \over (2\pi)^s} \ D_1(p-k) D_2(k),
\eeq
which is a convolution in momentum space,
corresponds in position space to the ordinary product
\beq
L(r) = D_1(r) D_2(r).
\eeq
This product is well-defined for long-range potentials such as $D_1(r) \sim r$.  This is the basic observation which gives a well-defined meaning to the DS equations for long-range forces.  

	One has the option of working entirely in position space.  However, as a matter of convenience, we may also work in momentum space, as usual, once we provide a well-defined two-way translation between position and momentum space.   

	The Fourier transform of a long-range correlator such as $D_1(r) = \sigma_{\rm coul} r$ is well-defined by providing a convergence factor, for example $\exp(-\epsilon r^2)$.  Indeed the integral
\beq
D(k) = \int d^3x \ \sigma_{\rm coul} r \ \exp( - i k \cdot x) \exp(-\epsilon r^2)  ,
\eeq
has a finite limit for $\epsilon \to 0$,
\beq
D(k) = - 8 \pi \sigma_{\rm coul} / k^4.
\eeq 
The minus sign here is as it should be because $\sigma_{\rm coul} r$ is an attractive potential.  [The minus sign in~(\ref{vcoul}) was in fact introduced to make $V_{\rm coul}(r)$ a positive quantity.]  We conclude that there is no difficulty in taking the Fourier transform of long-range potentials.

	However the inverse Fourier transform poses an apparent difficulty because it has an infrared divergence of the form $\int d^3k/k^4$.  Moreover the DS equations which contain a loop integral (\ref{convolution}) have infrared divergences of this type.  
	
	 To address these problems, consider the  inverse Fourier transform in $s$ spatial dimensions
\beq
\label{fourier}
\int {d^sk \over (2\pi)^s } \ { \exp(i k \cdot x) \over (k^2)^\alpha} = { \Gamma(s/2 - \alpha) \over 2^{2 \alpha } \ \pi^{s/2} \ \Gamma( \alpha) } \ r^{2 \alpha - s}.
\eeq	
It is free of infrared divergence provided that the parameter $\alpha$ satisfies the bound $\alpha < s/2$.  The integral has been evaluated by Gaussian integration after insertion of the identity,
\beq
{1 \over (k^2)^\alpha } = { 1 \over \Gamma(\alpha)} \int_0^\infty d\beta \ \beta^{\alpha - 1} \ \exp(- \beta k^2 ).
\eeq	
[By this method of integration the convergence factor $\exp(- \epsilon k^2)$ is not needed explicitly.]  The last integral is convergent only for $\alpha > 0$.  However the the left hand side of (\ref{fourier}) is well defined for all real $\alpha$ satisfying the bound $\alpha < s/2$ so, by analytic continuation, eq.~(\ref{fourier}) holds for all $\alpha$ that satisfies this bound.  Because $1/r^{s-2\alpha}$ appears on the right hand side of (\ref{fourier}), the restriction $\alpha < s/2$ corresponds to a negative power of $r$.  

	To extend the inverse Fourier transform to longer range potentials such as $r$ itself, which corresponds to a more infrared-singular power of $k$, we observe that such a more infrared-singular power may be written as a derivative,
\beq
\label{distribution}
{1 \over (k^2)^{1 + \alpha} } = 
{ -1 \over 4\alpha (s/2 - \alpha -1) } \ {\p^2 \over \p k_i^2} \ { 1 \over (k^2)^{\alpha} },
\eeq    
and we take the inverse Fourier transform of the right hand side in the distribution sense.  This gives
\beqa
\label{distribution2}
 && \int {d^sk \over (2\pi)^s } \  \exp(i k \cdot x) \  {1 \over (k^2)^{1 + \alpha} } 
\nonumber  \\
&=& { - 1  \over 4\alpha (s/2 - \alpha -1) }\int {d^sk \over (2\pi)^s } \ \exp(i k \cdot x)  \ {\p^2 \over \p k_i^2} \ { 1 \over (k^2)^{\alpha} }
\nonumber  \\
&=& { x^2  \over 4\alpha (s/2 - \alpha -1) }\int {d^sk \over (2\pi)^s } \ \exp(i k \cdot x) \ { 1 \over (k^2)^{\alpha} }
\nonumber  \\
&=& { \Gamma(s/2 - \alpha - 1) \over 2^{2 (\alpha + 1)} \ \pi^{s/2} \ \Gamma( \alpha + 1) } \ r^{2 (\alpha + 1) - s},
\eeqa	
where we have used (\ref{fourier}) for $\alpha < s/2$, and $r \equiv (x^2)^{1/2}$.  We now observe that (\ref{distribution2}) has the same form as (\ref{fourier}), with the substitution $\alpha \to \alpha + 1$.  Thus the result of giving $1/(k^2)^{\alpha + 1}$ a meaning as a distribution is quite simple: formula (\ref{fourier}) is continued analytically from $\alpha$ to $\alpha + 1$, so it is valid under the weaker restriction $\alpha < s/2 + 1$.  By induction, formula (\ref{fourier}) may be continued to $\alpha < s/2 + n$, where $n$ is any integer.

	Finally we note that the loop integral (\ref{convolution}) that appears in the DS equation becomes well-defined by the same method.  Indeed, suppose that $D_1(k) = 1/(k^2)^{\alpha + 1}$ is written as the derivative 
\beq	
D_1(k) = { \p^2 \over \p k_i^2 } \ E_1(k)
\eeq 
where $E_1(k)$ is written above and is less singular than $D_1(k)$.  Then the ill-defined loop integral (\ref{convolution}) may be replaced by the well-defined expression
\beq
\label{convolution2}
L(p) =  { \p^2 \over \p p_i^2 }\int {d^sk \over (2\pi)^s} \ E_1(p-k) D_2(k).
\eeq
However there is no need to do this explicitly because the result of this substitution may be obtained, as before, by analytic continuation in $\alpha$ of the original loop integral $\int {d^sk \over (2\pi)^s} \ D_1(p-k) D_2(k)$.  

	We conclude that instead of working in position space, where there are no infrared divergences, we may, as a matter of convenience, work directly in momentum space according to the following prescription: The standard loop integral (\ref{convolution}) is performed for values of the critical exponents $\alpha$ for which it is well defined.  The result is then analytically continued in $\alpha$ to the values of interest.

\section{Definition of infrared critical exponents}
\label{sec:IRExpFinT_IRDef}

The only dimensionful parameter in the DSEs is $g^2T$,
which in spatial dimension $s$ provides a mass scale $m$
defined by
\beq
  m^{4-s} = g^2 T.
  \label{eq:massscale}
\eeq
As an ansatz we look for a solution to the DSE for the one-Matsubara frequency propagator
that is a simple power law in the spatial momentum,
\beqa
D_{\bf AA}(k) \sim {b_{\bf A} m^{2\alpha_{\bf A} } \over (k^2)^{1 + \alpha_{\bf A} } }; 
\ \ \ \ \ \
D_{\pi' \pi'}(k) \sim {b_{\pi'} m^{2\alpha_{\pi'} } \over (k^2)^{\alpha_{\pi'} } };
\nonumber  \\
D_{c \bar c}(k) \sim {b_{\rm gh}m^{2\alpha_{\rm gh} } \over (k^2)^{1 + \alpha_{\rm gh} } };  \ \ \ \
D_{A_0 A_0}(k) \sim {b_0 m^{2\alpha_0 } \over (k^2)^{1 + \alpha_0 } };
\label{eq:powerlawansatz}  \\
D_{\varphi A_0}(k) \sim {- \I b_m m^{2\alpha_m } \over (k^2)^{1 + \alpha_m } };
 \ \ \ \ \ \
D_{\varphi \varphi}(k) \sim {b_\varphi m^{2\alpha_{\varphi} } \over (k^2)^{1 + \alpha_\varphi } }\,. \nonumber
\eeqa
In the following it will be sometimes convenient to write $\alpha_{\pi'}=1+\hat\alpha_{\pi'}$. 
The mass $m$ cancels out of all equations because of engineering dimensions.

\section{Infrared asymptotic DS equations}
\label{sec:IRExpFinT_IRDSEs}
	
In the DS equations, we take the external momentum $k$ and the loop momentum
$p$ to be small compared to the other scales in the theory, and we take the infrared
asymptotic form of the propagators.  This will yield a finite system of equations which
will provide a self-consistent infrared limit of the propagators.
	
The infrared asymptotic equations read (with loop integrals $I_{S(V, S)}$,
$I_{S(V, V)}$ etc. to be defined in subsection~\ref{ssec:IRExp_defloop})
for $\Gamma_{A_0 A_0}$\footnote{We have checked explicitly
that in the infrared, given the DS equations of Figure~\ref{fig:IRCoulomb_firstDSEs},
both the assumption $D_{\varphi \varphi} D_{A_0 A_0} \ll D_{A_0 \varphi}^2$ and
$D_{\varphi \varphi} D_{A_0 A_0} \gg D_{A_0 \varphi}^2$ lead to a contradiction
for this system of equations. Thus $D_{\varphi \varphi}D_{A_0 A_0}$ and
$D_{A_0 \varphi}^2$ have the same infrared behavior. A possibility we have
not further explored in this article is a cancellation $b_0b_{\varphi}+b_m^2=0$.}
\begin{align}
{b_\varphi \over b_0 b_\varphi + b_m^2} 
&= b_{\bf A} b_\varphi I_{S(V, S)} (\alpha_{\bf A}, \alpha_\varphi) \nonumber \\
&\quad+b_{\bf A} b_{\pi'} I_{S(V, V)} (\alpha_{\bf A}, \hat\alpha_{\pi'});
\label{GA0A0}	
\end{align}
for $\Gamma_{\varphi A_0}$,
\begin{equation}
\label{GphiA0}	
{b_m \over b_0 b_\varphi + b_m^2} 
= - b_{\bf A} b_m I_{S(V, S)} (\alpha_{\bf A}, \alpha_m);
\end{equation}
for $\Gamma_{\varphi \varphi}$,
\begin{equation}
\label{Gphiphi}	
{b_0 \over b_0 b_\varphi + b_m^2} 
= b_{\bf A} b_0 I_{S(V, S)} (\alpha_{\bf A}, \alpha_0);
\end{equation}
for $\Gamma_{\bf AA}$,
\begin{eqnarray}
\label{GAA}	
1  & = &  b_{\bf A} b_0 b_{\pi'} I_{V(V, S)} (\hat\alpha_{\pi'}, \alpha_0)
+ b_{\bf A} b_0 b_\varphi I_{V(S, S)} (\alpha_0, \alpha_\varphi)
\nonumber  \\  
&&- b_{\bf A} b_m^2 I_{V(S, S)} (\alpha_m, \alpha_m)
+ b_{\bf A} b_{\rm gh}^2 I_{V(S, S)} (\alpha_{\rm gh}, \alpha_{\rm gh}).
\nonumber \\    \ 
\end{eqnarray}
for $\Gamma_{\pi' \pi'}$,
\begin{equation}
\label{Gpipi}	
1 = b_{\bf A} b_0 b_{\pi'} I_{V(V, S)} (\alpha_{\bf A}, \alpha_0);
\end{equation}
for $\Gamma_{\bar c c}$,
\begin{equation}
\label{Gghgh}	
1 = - b_{\bf A} b_{\rm gh}^2 I_{S(V, S)} (\alpha_{\bf A}, \alpha_{\rm gh})\,.
\end{equation}

\subsection{Symmetry of the infrared asymptotic equations}

There is a two-parameter continuous symmetry transformation that
these equations inherit from the cubic interaction terms 
$\pi_i gA_i \times A_0$ and $\p_i \bar c g A_i \times c$, namely
\beqa
A_i & \to & \exp(i\beta)A_i; \ \ \ \
A_0 \to \exp(i\gamma)A_0; \ \ \ \
c \to \exp(i\gamma)c 
\nonumber    \\
\pi_i & \to & \exp[-i(\beta + \gamma)] \pi_i; \ \ \ \ \ 
\bar c \to \exp[-i(\beta + \gamma)] \bar c.
\eeqa
As a consequence of this symmetry, the infrared asymptotic DS equations
are invariant under the transformations of the asymptotic propagators
\beq
  \begin{array}{lcl}
  \label{symofbs}
b_{\bf A} \to \exp(2\I\beta) b_{\bf A}; &&
b_{\pi'} \to \exp[-2\I(\beta + \gamma)] \ b_{\pi'} \\[6pt]
b_{\rm gh} \to \exp(-\I \beta) b_{\rm gh}; &&
b_0 \to \exp(2\I\gamma) \ b_0; \\[6pt]
b_m \to \exp(-\I \beta)b_m; &&
b_\varphi \to \exp[-2\I(\beta + \gamma)] \ b_\varphi.
  \end{array}
\eeq
Because of this 2-parameter symmetry the DS equations provide
only 4 relations among the 6 $b$-coefficients.

\subsection{Definition of loop integrals}
\label{ssec:IRExp_defloop}

We now define the symbols that represent the loop integrals,
\beqa
\label{ISVS}
I_{S(V,S)}(\alpha_{\bf A}, \alpha_\varphi) &\equiv&
N k^{-s +2 \alpha_{\bf A} + 2 \alpha_\varphi + 2} 
\\   \nonumber  
&&\times \int { d^s p \over (2 \pi)^s } \ 
{k^2 p^2 - (p \cdot k)^2 \over (p^2)^{2+\alpha_{\bf A}} \ [(k - p)^{2}]^{1 + \alpha_\varphi}}
\eeqa
\beqa
\label{IVVS}
I_{V(V,S)}(\alpha_{\bf A}, \alpha_0) & \equiv &
{N k^{-s +2 \alpha_{\bf A} + 2 \alpha_0 + 2} \over s-1 } 
\\   \nonumber
&&  \times \int { d^s p \over (2 \pi)^s } \ 
{(s-2)k^2 p^2 + (p \cdot k)^2 \over 
(p^2)^{2+\alpha_{\bf A}} \ [(k - p)^{2}]^{1 + \alpha_0}}
\eeqa
\beqa
\label{ISVV}
I_{S(V,V)}(\alpha_{\bf A}, \hat\alpha_{\pi'}) & \equiv &
N k^{-s +2 \alpha_{\bf A} + 2 \hat\alpha_{\pi'} + 4} 
\int { d^s p \over (2 \pi)^s }
\\   \nonumber
&& \times 
{(s-2)(k-p)^2 p^2 + [(k-p) \cdot p]^2 \over (p^2)^{2+\alpha_{\bf A}} \ [(k - p)^{2}]^{2 + \hat\alpha_{\pi'}}}.
\eeqa
\beqa
\label{IVSS}
I_{V(S,S)}(\alpha_0, \alpha_\varphi) & \equiv &
{N k^{-s +2 \alpha_0 + 2 \alpha_\varphi} \over s-1 }
\\   \nonumber
&& \times \int { d^s p \over (2 \pi)^s } \ 
{k^2 p^2 - (p \cdot k)^2 \over (p^2)^{1+\alpha_0} \ [(k - p)^{2}]^{1 + \alpha_\varphi}}.
\eeqa
One sees by inspection that two of the symbols are related by
\begin{equation}
\label{VSSSVS}
I_{V(S,S)}(\alpha_0, \alpha_\varphi) = (s-1)^{-1} I_{S(V,S)}(\alpha_0 -1, \alpha_\varphi).
\end{equation}

The symbols have the value (see appendix~\ref{app:IRExpFinT_PowerLaw}
for the exemplary evaluation of one such integral)
\beqa
\label{valISVS}
I_{S(V,S)}(\alpha_{\bf A}, \alpha_{\rm gh}) = 
{N (s-1) \over 2 (4 \pi)^{s/2}} \  
{ \Gamma(2 + \alpha_{\bf A} + \alpha_{\rm gh} - s/2)  
 \over 
 \Gamma(2 + \alpha_{\bf A})  }
\nonumber  \\
\times {  \Gamma(s/2 - \alpha_{\rm gh})  \ \Gamma(s/2 - \alpha_{\bf A} -1) 
 \over 
 \Gamma(1 + \alpha_{\rm gh})  \ \Gamma(s - \alpha_{\rm gh} - \alpha_{\bf A} -1) },
 \nonumber   \\   \
\eeqa
\begin{eqnarray}
\label{valIVVS}
I_{V(V,S)}(\alpha_{\bf A}, \alpha_0) & = & {N c_1 \over 4 (4 \pi)^{s/2}} 
{ \Gamma(2 + \alpha_{\bf A} + \alpha_0 - s/2) 
 \over 
 \Gamma(2 + \alpha_{\bf A})  }
 \\ \nonumber
&& \times { \Gamma(s/2 - \alpha_0 -1)  \ \Gamma(s/2 - \alpha_{\bf A} -1) 
 \over 
 \Gamma(1 + \alpha_0)  \ \Gamma(s - \alpha_0 - \alpha_{\bf A} -1) },
\end{eqnarray}
where
\beq
c_1 \equiv   (s-1)( 4 \alpha_{\bf A} + 3)
- (2\alpha_{\bf A} + 2\alpha_0 + 3)(2\alpha_{\bf A} + 1),
\eeq
\begin{eqnarray}
\label{valISVV}
I_{S(V,V)}(\alpha_{\bf A}, \hat\alpha_{\pi'}) & = & {N c_2 \over 4(4 \pi)^{s/2}} 
{ \Gamma(2 + \alpha_{\bf A} + \hat\alpha_{\pi'} - s/2)  
 \over 
 \Gamma(2 + \alpha_{\bf A}) }
\nonumber   \\ 
&  \times & { \Gamma(s/2 - \hat\alpha_{\pi'} -1)  \ \Gamma(s/2 - \alpha_{\bf A} -1) 
 \over 
 \Gamma(2 + \hat\alpha_{\pi'})  \ \Gamma(s - \alpha_{\bf A} - \hat\alpha_{\pi'} -2 ) }
 \nonumber \\  \
\end{eqnarray}
where
\beq
c_2 \equiv (s-1) \ [s -1 +( 1 + 2 \alpha_{\bf A})(1 + 2 \hat\alpha_{\pi'}) ]  
\eeq
\beqa
\label{valIVSS}
I_{V(S,S)}(\alpha_0, \alpha_\varphi) = 
{N \over 2 (4 \pi)^{s/2}} \  
{ \Gamma(1 + \alpha_0 + \alpha_\varphi - s/2) \  
 \over 
 \Gamma(1 + \alpha_0)  }
 \nonumber  \\
 \times {  \Gamma(s/2 - \alpha_\varphi)  \ \Gamma(s/2 - \alpha_0) 
 \over 
 \Gamma(1 + \alpha_\varphi)  \ \Gamma(s - \alpha_\varphi - \alpha_0) }.
\eeqa

\subsection{Check of loop integrals}

We obtain a useful check on the evaluation of the loop integrals by
rewriting\footnote{In the following the expression $p_1p_2$ denotes
the dyadic product of the $s$-vectors $p_1$ and $p_2$. In a less
compact way, one would write~\eqref{eq:IRExp_rewritenum} as
\[
  \mathcal{N} \equiv {\rm Tr}_s \left[ \left(\norm{\V{q}}^2\1_s - \V{q}\,\V{q}^{\top}\right)
     \left(\norm{\V{p}}^2\1_s - \V{p}\,\V{p}^{\top}\right) \right]\,,
\]
where $\V{p}$ and $\V{q}$ are column vectors and $\top$ denotes
transposition.} the numerator of the integrand of $I_{S(V,V)}$,
\beq
  \mathcal{N} \equiv {\rm Tr} [ (q^2 - qq)(p^2 - pp) ],
  \label{eq:IRExp_rewritenum}
\eeq
where $q = k-p$.  We have
\beqa
T & = & q^2 (s-1) p^2 - q \cdot (p^2 - pp) \cdot q
\nonumber \\
& = & {q^2 \over k^2} \  k^2 (s -1) p^2 - k \cdot (p^2 - pp) \cdot k 
\nonumber  \\ & = & 
{q^2 \over k^2} \ {\rm Tr} [ (k^2 - kk)(p^2 - pp) ]
\nonumber  \\
&& + \Big( {q^2 \over k^2} -1 \Big)
k \cdot (p^2 - pp) \cdot k,
\eeqa
which, by comparison with (\ref{ISVS}) through (\ref{IVSS}), leads to the identity 
\beqa
I_{S(V,V)}(\alpha_{\bf A}, \hat\alpha_{\pi'}) 
& = & (s-1) I_{V(V,S)}(\alpha_{\bf A}, \hat\alpha_{\pi'}) 
\nonumber   \\
&& + I_{S(V,S)}(\alpha_{\bf A}, \hat\alpha_{\pi'})
\\   \nonumber 
&& - (s-1)I_{V(S,S)}(1 + \alpha_{\bf A}, 1+ \hat\alpha_{\pi'}).
\eeqa
As a precise check, it has been verified that this relation between the 4 integrals is satisfied by the 4 values just given.

\section{Determination of infrared critical exponents} 
\label{sec:IRExpFinT_DetCritExp}

\subsection{4 power-relations among infrared critical exponents}

There are 6 infrared critical exponents and 6 DS equations. 
By equating powers of momentum on both sides of the DS equations,
one obtains relations between the infrared critical exponents.
From the equation for $\Gamma_{\bar c c}$, one obtains
\begin{equation}
  2 \alpha_{\rm gh} + \alpha_{\bf A} = (s - 4)/2;
  \label{eq:relation_alphagh_alphaA}
\end{equation}
from the equation for $\Gamma_{\varphi \varphi}$,
\begin{equation}
\alpha_0 + \alpha_\varphi + \alpha_{\bf A} = (s - 4)/2;
\end{equation}
from the equation for $\Gamma_{\varphi A_0}$,
\begin{equation}
2 \alpha_m + \alpha_{\bf A} = (s - 4)/2;
\end{equation}
from the equation for $\Gamma_{\pi' \pi'}$,
\begin{equation}
\alpha_0 + \alpha_{\pi'} + \alpha_{\bf A} = (s - 4)/2.
\end{equation}
These 4 power relations come from the 4 DS equations that have only one term on the right-hand side.  They leave undetermined two infrared critical exponents which we may choose to be $\alpha_{\rm gh}$ and~$\alpha_0$.  The remaining infrared critical exponents may be expressed in terms of these by
\begin{align}
  \alpha_{\bf A} &= (s - 4)/2 - 2 \alpha_{\rm gh}\,, \nonumber \\
  \alpha_m &= \alpha_{\rm gh}\,,  \label{eq:IRExp_powercountrel} \\  
  \alpha_{\pi'} &= \alpha_\varphi = 2 \alpha_{\rm gh} - \alpha_0\,. \nonumber
\end{align}

The equation
\begin{equation}
\alpha_0 + \alpha_\varphi = 2 \alpha_m,
\end{equation}
which follows from the above, relates the critical exponents of
$D_{A_0A_0} \sim 1/k^{2 + 2\alpha_0}$,
$D_{\varphi \varphi} \sim 1/k^{2 + 2\alpha_\varphi}$ and
$D_{\varphi \alpha_0}\sim 1/k^{2 + \alpha_\varphi + \alpha_0}$.
Thus the infrared critical exponents $\alpha_0$ and $\alpha_\varphi$
characterize the elementary fields $A_0$ and $\varphi$.

\subsection{Equations for $\alpha_0$ and $\alpha_{\rm gh}$}

When the above 4~power relations on the 6 critical exponents (the 6~$\alpha$'s) are satisfied, the power relations among the remaining 2~DS equations are satisfied identically.  The 6~DS equations also provide 6~relations among the 6 $b$-coefficients.  However, because of the 2-parameter symmetry invariance~(\ref{symofbs}), of these~6 equations, only~4 are independent conditions on the $b$-coefficients. The remaining two equations provide consistency conditions that determine the two missing relations among the infrared critical exponents, as we now show.  Thus all~6 infrared critical exponents are determined. 

From (\ref{GphiA0}) and (\ref{Gphiphi}) we obtain an equation relating critical exponents,
\begin{equation}
\label{Gamma1}
I_{S(V, S)} (\alpha_{\bf A}, \alpha_{\rm gh}) = - I_{S(V, S)} (\alpha_{\bf A}, \alpha_0),
\end{equation}
where we have used $\alpha_{\rm gh} =  \alpha_m$.

The remaining relation between critical exponents is obtained as follows.  From (\ref{GphiA0}) and (\ref{Gghgh}) and power relation $\alpha_{\rm gh} = \alpha_m$ we obtain
\begin{equation}
\label{bghbm}
b_{\rm gh}^2 = b_0 b_\varphi + b_m^2. 
\end{equation}
This equation allows us to write (\ref{GA0A0}) as
\begin{equation}
b_\varphi = 
b_{\bf A} b_\varphi b_{\rm gh}^2 I_{S(V,S)}(\alpha_{\bf A}, \alpha_\varphi)
+ b_{\bf A} b_{\pi'} b_{\rm gh}^2 I_{S(V,V)}(\alpha_{\bf A}, \hat\alpha_{\pi'})
\end{equation}	
or, by (\ref{Gghgh}),
\begin{equation}
\label{F1}
{b_{\pi'} \over b_\varphi} = {- I_{S(V,S)}(\alpha_{\bf A}, \alpha_{\rm gh}) - I_{S(V,S)}(\alpha_{\bf A}, \alpha_\varphi) \over I_{S(V,V)}(\alpha_{\bf A}, \hat\alpha_{\pi'}) }  \equiv F_1.
\end{equation}

Likewise (\ref{bghbm}) allows us to write (\ref{GAA}) as
\beqa
1 & = & b_{\bf A} b_0 b_{\pi'} I_{V(V,S)}(\hat\alpha_{\pi'}, \alpha_0) +
b_{\bf A} b_0 b_\varphi I_{V(S,S)}(\alpha_0, \alpha_\varphi)
\nonumber   \\ 
&& + b_{\bf A} ( b_0 b_\varphi - b_{\rm gh}^2) I_{V(S,S)}(\alpha_m, \alpha_m)
\nonumber  \\
&& + b_{\bf A}  b_{\rm gh}^2 I_{V(S,S)}(\alpha_{\rm gh}, \alpha_{\rm gh}).
\eeqa
With $\alpha_m = \alpha_{\rm gh}$, there is a partial cancellation between the last two terms which are the contribution from bose and fermi ghost loops respectively, and we obtain
\beqa
1 & = & b_{\bf A} b_0 b_{\pi'} I_{V(V,S)}(\hat\alpha_{\pi'}, \alpha_0) 
\\   \nonumber 
&&+ b_{\bf A} b_0 b_\varphi [ I_{V(S,S)}(\alpha_0, \alpha_\varphi)
+ I_{V(S,S)}(\alpha_{\rm gh}, \alpha_{\rm gh})].
\eeqa
This gives, by (\ref{Gpipi})
\begin{equation}
\label{F2}
{b_{\pi'} \over b_\varphi} =  
{  I_{V(S,S)}(\alpha_0, \alpha_\varphi)
+ I_{V(S,S)}(\alpha_{\rm gh}, \alpha_{\rm gh})
\over 
I_{V(V,S)}(\alpha_{\bf A}, \alpha_0)
- I_{V(V,S)}(\hat\alpha_{\pi'}, \alpha_0) } \equiv F_2.
\end{equation}
From equations (\ref{F1}) and (\ref{F2}) we obtain the final equation that determines the infrared critical exponents.
\begin{equation}
  \label{Gamma2}
  F_1 - F_2 = 0.
\end{equation}

Equations~\eqref{Gamma1} and~\eqref{Gamma2} together with the
above power relations given previously determine the remaining two
infrared critical exponents, $\alpha_0$ and $\alpha_{\rm gh}$.

\subsection{General Remarks on the Equations}

The driving force of the system seems to be equation (\ref{Gamma1}).
Upon canceling common factors, this equation reads, from (\ref{valISVS}),
\beqa
&& { - \Gamma(2 + \alpha_{\bf A} + \alpha_{\rm gh} - s/2) \ \Gamma(s/2 - \alpha_{\rm gh})   \over 
 \Gamma(1 + \alpha_{\rm gh})  \ \Gamma(s - \alpha_{\rm gh} - \alpha_{\bf A} -1) }
\\   \nonumber  
&& \ \ \ \ \ \ \ \ \ \ 
= { \Gamma(2 + \alpha_{\bf A} + \alpha_0 - s/2) \ \Gamma(s/2 - \alpha_0)   \over 
 \Gamma(1 + \alpha_0)  \ \Gamma(s - \alpha_0 - \alpha_{\bf A} -1) }.
\eeqa 
We  eliminate $\alpha_{\bf A}$ using
$\alpha_{\bf A} = -2 \alpha_{\rm gh} + (s - 4)/2$ and obtain
\beqa
\label{valGamma1}
&&  { - \Gamma(- \alpha_{\rm gh}) \ \Gamma(s/2 - \alpha_{\rm gh})   \over 
 \Gamma(1 + \alpha_{\rm gh})  \ \Gamma((s+2)/2 + \alpha_{\rm gh}) }
 \\   \nonumber  
&& \ \ \ \ \ \ \ \ \ \ 
 = 
 { \Gamma(\alpha_0 - 2 \alpha_{\rm gh}) \ \Gamma(s/2 - \alpha_0)   \over 
 \Gamma(1 + \alpha_0)  \ \Gamma((s+2)/2 + 2\alpha_{\rm gh} - \alpha_0) }.
\eeqa
Note that
\begin{equation}
\label{Gammaid}
- \Gamma(- \alpha_{\rm gh}) = 
{ \Gamma(1-\alpha_{\rm gh}) \over \alpha_{\rm gh} },
\end{equation}
so this factor is positive for $0 < \alpha_{\rm gh} <1$.

Now let $s$ be fixed in the interval
\begin{equation}
1 < s \le 3.
\end{equation}
and let $\alpha_0$ be fixed in the interval
\begin{equation}
0 < \alpha_0 < s/2 \le 3/2.
\end{equation}
Then for $\alpha_{\rm gh}$ in the interval
\begin{equation}
0 < \alpha_{\rm gh} < \alpha_0/2 \le 3/4
\end{equation}
both sides of (\ref{valGamma1}) are positive and finite.  Moreover, by (\ref{Gammaid}), when $\alpha_{\rm gh}$ approaches its lower limit, namely 0, the LHS of (\ref{valGamma1}) approaches $+ \infty$ while the RHS is finite, and when $\alpha_{\rm gh}$ approaches its upper limit, namely $\alpha_0/2$, the RHS approaches $+ \infty$ while the LHS remains finite.  Since there are no further poles or other discontinuities present in the stated interval, for every $\alpha_0\in(0,s/2)$ there exists (at least) one solution $\alpha_{\rm gh}$. 

This tells us that to solve (\ref{valGamma1}) numerically we should take $\alpha_0$ as the independent variable and we are assured that there exists a solution for
\begin{equation}
  \alpha_{\rm gh} = \alpha_{\rm gh}(\alpha_0),
\end{equation}
both variables being in the stated intervals.

\subsection{Analytic Statements}

We may in fact solve (\ref{valGamma1}) analytically for $\alpha_0$ close to its end-points, $\alpha_0 = \epsilon$ and $\alpha_0 = s/2 - \epsilon$, where $\epsilon$ is small.  Suppose first that $\alpha_0 = \epsilon$.  Then the inequality $0 < \alpha_{\rm gh} < \epsilon/2$ implies that $\alpha_{\rm gh}$ is also small.  In this limit (\ref{valGamma1}) approaches
\begin{equation}
{ - \Gamma(- \alpha_{\rm gh}) \ \Gamma(s/2)   \over 
 \Gamma(1)  \ \Gamma((s+2)/2) }
 = 
 { \Gamma(\epsilon - 2 \alpha_{\rm gh}) \ \Gamma(s/2)   \over 
 \Gamma(1)  \ \Gamma((s+2)/2) }.
\end{equation}
With $- \Gamma(- \alpha_{\rm gh}) \approx 1/ \alpha_{\rm gh}$, and $\Gamma(\epsilon - 2 \alpha_{\rm gh}) \approx 1/ (\epsilon - 2\alpha_{\rm gh})$, we may equate the singular pole terms
\begin{equation} 
 {1 \over \alpha_{\rm gh} } =  {1 \over \epsilon - 2\alpha_{\rm gh} },
\end{equation}
and with $\epsilon = \alpha_0$, this has the solution
\begin{equation}  
\label{smallalpha0}
 \alpha_{\rm gh}(\alpha_0) = \alpha_0/3  \ \ \ \ \ \ \ \ \ \ \ \ \ \ \ \ \ 
 \alpha_0 \approx 0.
\end{equation}
 Now suppose that $\alpha_0 = s/2 - \epsilon$.  Then since the RHS of (\ref{valGamma1}) blows up like $1/ \epsilon$, $\alpha_{\rm gh}$ must be small, so (\ref{valGamma1}) approaches
\begin{equation}
{ - \Gamma(- \alpha_{\rm gh}) \ \Gamma(s/2)   \over 
 \Gamma(1)  \ \Gamma((s+2)/2) }
 = 
 { \Gamma(s/2) \ \Gamma(\epsilon)   \over 
 \Gamma(1 + s/2)  \ \Gamma(1) }.
\end{equation}
 We again equate singular pole terms
\begin{equation}
{1 \over \alpha_{\rm gh} } = {1 \over \epsilon }
\end{equation}
 which, with $\epsilon = s/2 - \alpha_0$, gives
\begin{equation}
 \alpha_{\rm gh}(\alpha_0) = s/2 - \alpha_0  
  \ \ \ \ \ \ \ \ \ \ \ 
 \alpha_0 \approx s/2 - \epsilon.
 \end{equation}
We have now determined that $\alpha_{\rm gh}(\alpha_0)$ vanishes at $\alpha_0 = 0, s/2$, and we have determined its slope at these two points.  We approximate $\alpha_{\rm gh}(\alpha_0)$ in its interval by an interpolation.  We write
\begin{equation}
\alpha_{\rm gh} = \alpha_0 (s/2 - \alpha_0) f(\alpha_0),
\end{equation}
where $f(\alpha_0)$ is a positive function whose values at $\alpha_0 = 0$ and $\alpha_0 = s/2$ are determined by the slope of $\alpha_{\rm gh}(\alpha_0)$ at these two points which are given respectively by $1/3$ and $-1$.

A linear interpolation for $f(\alpha_0)$ yields an
approximate solution to (\ref{valGamma1}),
\begin{equation}
\alpha_{\rm gh}(\alpha_0) \approx \alpha_0 \left(\frac{s}2 - \alpha_0\right) {2 \over 3s }
\Big(1 + {4 \alpha_0 \over s}\Big)\,,
\end{equation}
and comparison with the numerical results
in subsection~\ref{ssec:IRExp_numres} shows
that this is a reasonable approximation (with less
than 10\% deviation) even for intermediate
values of $\alpha_0$.

\subsection{Numerical Results}
\label{ssec:IRExp_numres}

A full analytic solution of~\eqref{Gamma1} and~\eqref{Gamma2} has not
been obtained so far. To find a solution at least numerically, one can
follow one of two strategies:
\begin{itemize}
\item One equation (preferably~\eqref{Gamma1}) can be solved
  for one variable (yielding $\alpha_{\rm gh}^{(1)}(\alpha_0)$)  and this
  function can be substituted into both $F_1$ from~\eqref{F1} and
  $F_2$ from~\eqref{F2}, yielding two functions of one variable
  $F_i(\alpha_0,\,\alpha_{\rm gh}(\alpha_0))$, $i=1,2$. Intersection
  points of these functions are solutions of the system of equations.
\item Equations~\eqref{Gamma1} and~\eqref{Gamma2} each
  implictly define a (possibly multi-valued) function
  $\alpha_{\rm gh}^{(i)}(\alpha_0)$, $i=1,2$. One can obtain each
  of these functions separately and find the solutions of the system
  as intersection points in a two-dimensional plot.
\end{itemize}

The second strategy is more cumbersome, yet it gives a better
understanding of the conditioning of the system and the relationship
between solutions for various values of $s$. While a solution
for~\eqref{Gamma1} is easy to find, solving~\eqref{Gamma2}
requires more effort, since it defines a multivalued function with
several potentially relevant branches.

\begin{figure*}
    \includegraphics[width=18cm]{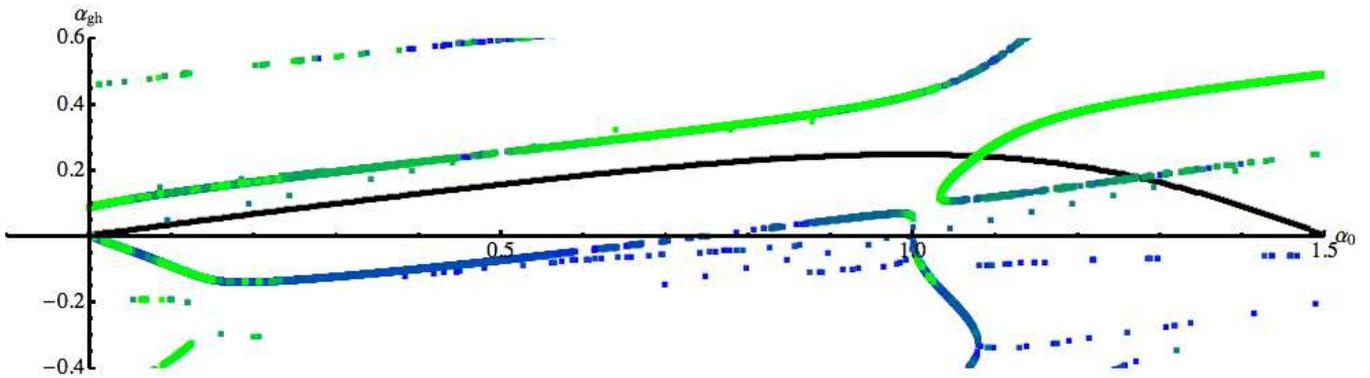}
  \caption{Solution of equations~\eqref{Gamma1}
    and~\eqref{Gamma2} for $s=3$: We plot $\alpha_{\rm gh}(\alpha_0)$
    from~\eqref{Gamma1} and from~\eqref{Gamma2}. The relevant solution
    of the first first equation is represented by a solid black line.
    The second equation gives rise to a multivalued function: The graphs
    are composed of single dots; different shades of blue and green
    (in the color version) indicate different initial guesses. Note that
    single points which do not belong to any branch of the function
    typically correspond to values in which the \texttt{findroot}
    routine got stuck.}
  \label{fig:IRExp_seq3}
\end{figure*}

\begin{figure*}
    \includegraphics[width=17cm]{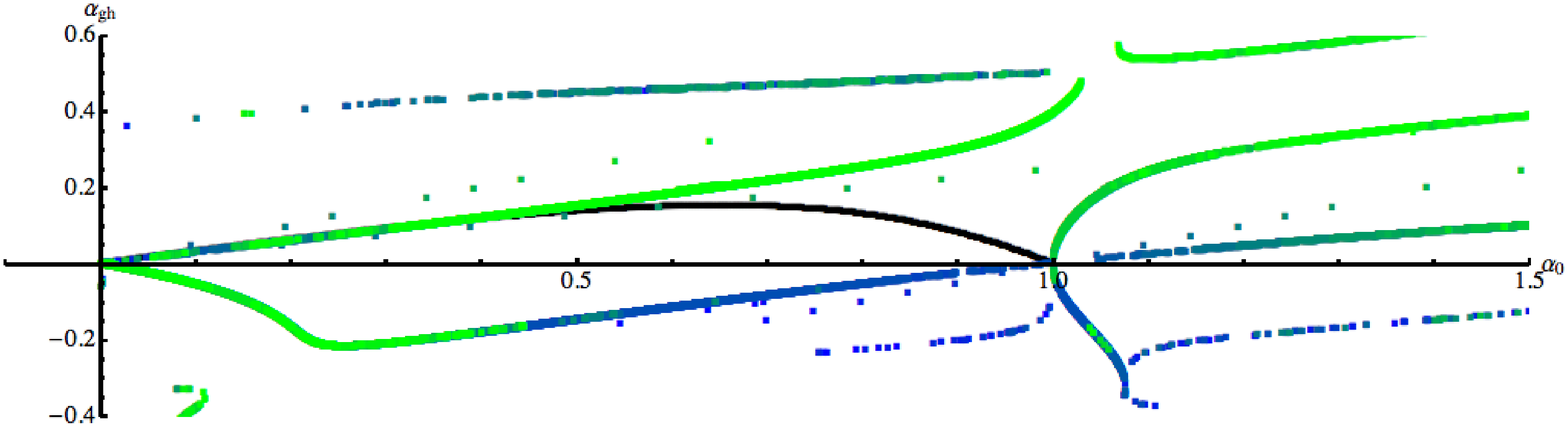}
  \caption{Solution of equations~\eqref{Gamma1}
    and~\eqref{Gamma2} for $s=2$, otherwise as in Figure~\ref{fig:IRExp_seq3}.}
  \label{fig:IRExp_seq2}
\end{figure*}

\begin{figure*}
    \includegraphics[width=17cm]{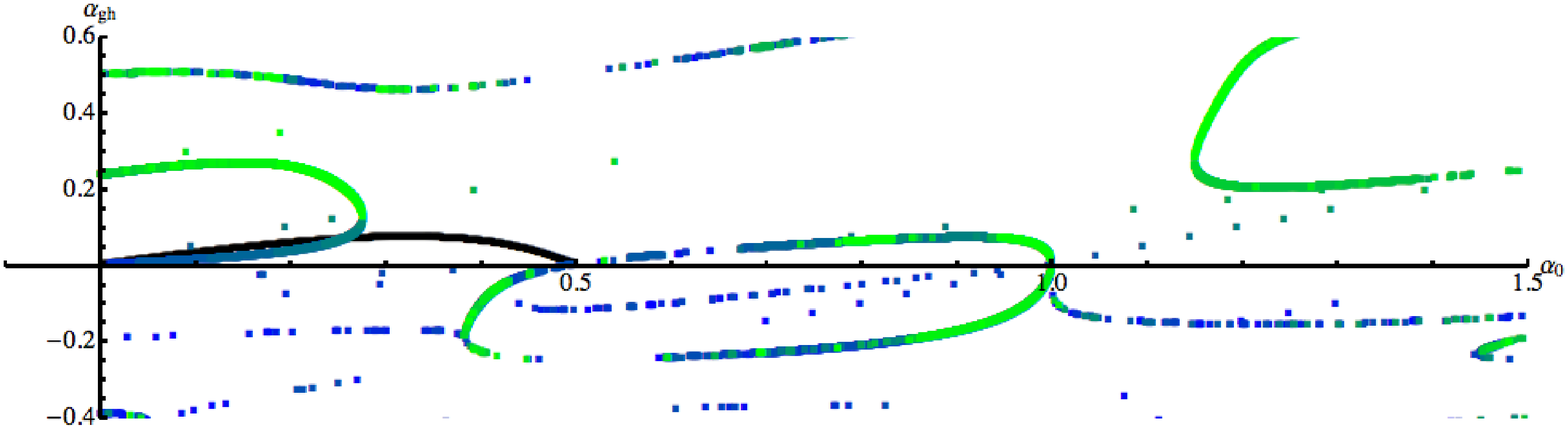}
  \caption{Solution of equations~\eqref{Gamma1}
    and~\eqref{Gamma2} for $s=1$, otherwise as in Figure~\ref{fig:IRExp_seq3}.}
  \label{fig:IRExp_seq1}
\end{figure*}

\begin{figure}
    \includegraphics[width=8.5cm]{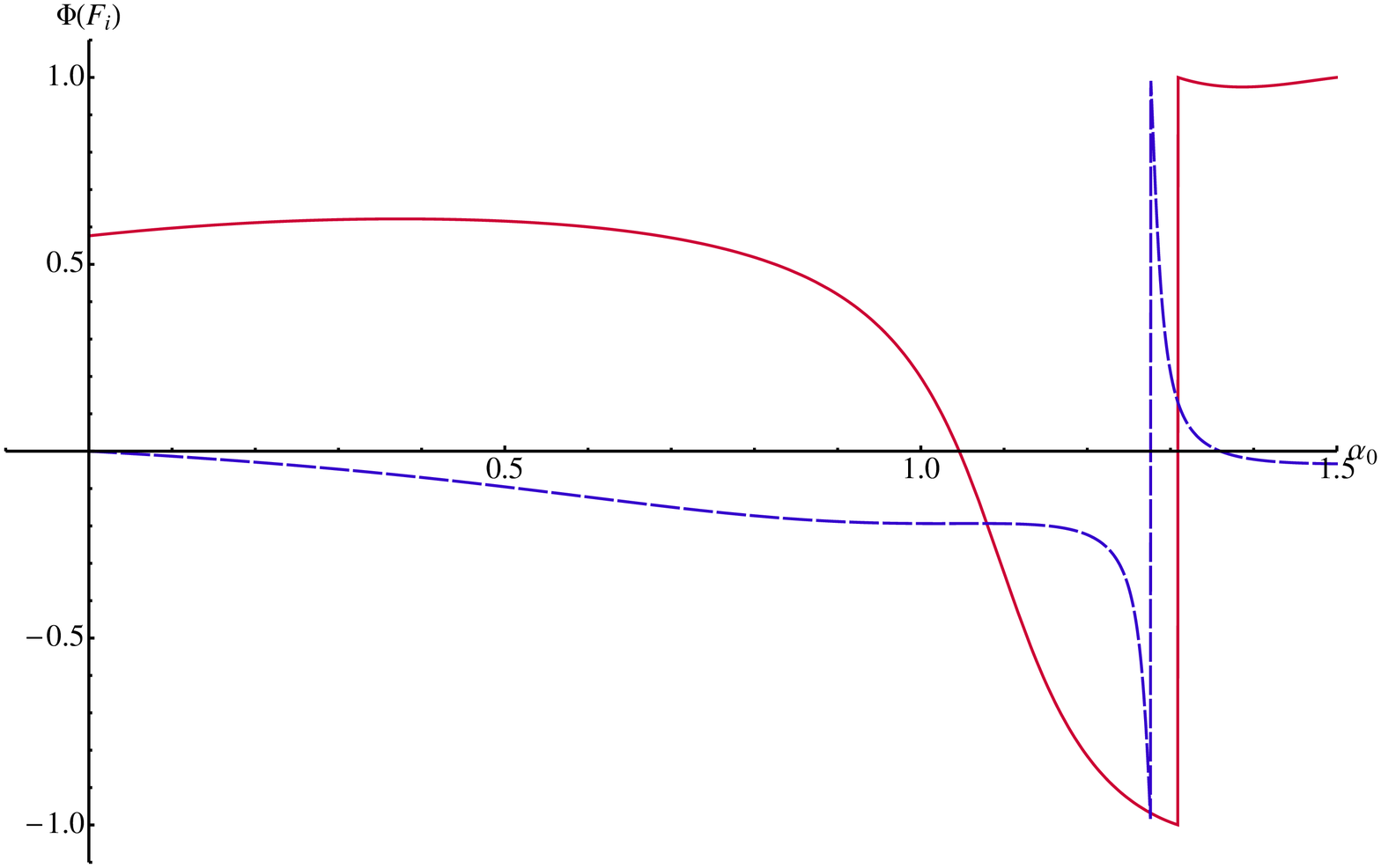} \put(-240,160){(a)\hspace{1cm}\raisebox{-12pt}{$s=3$}} \\[12pt]
    \includegraphics[width=8.5cm]{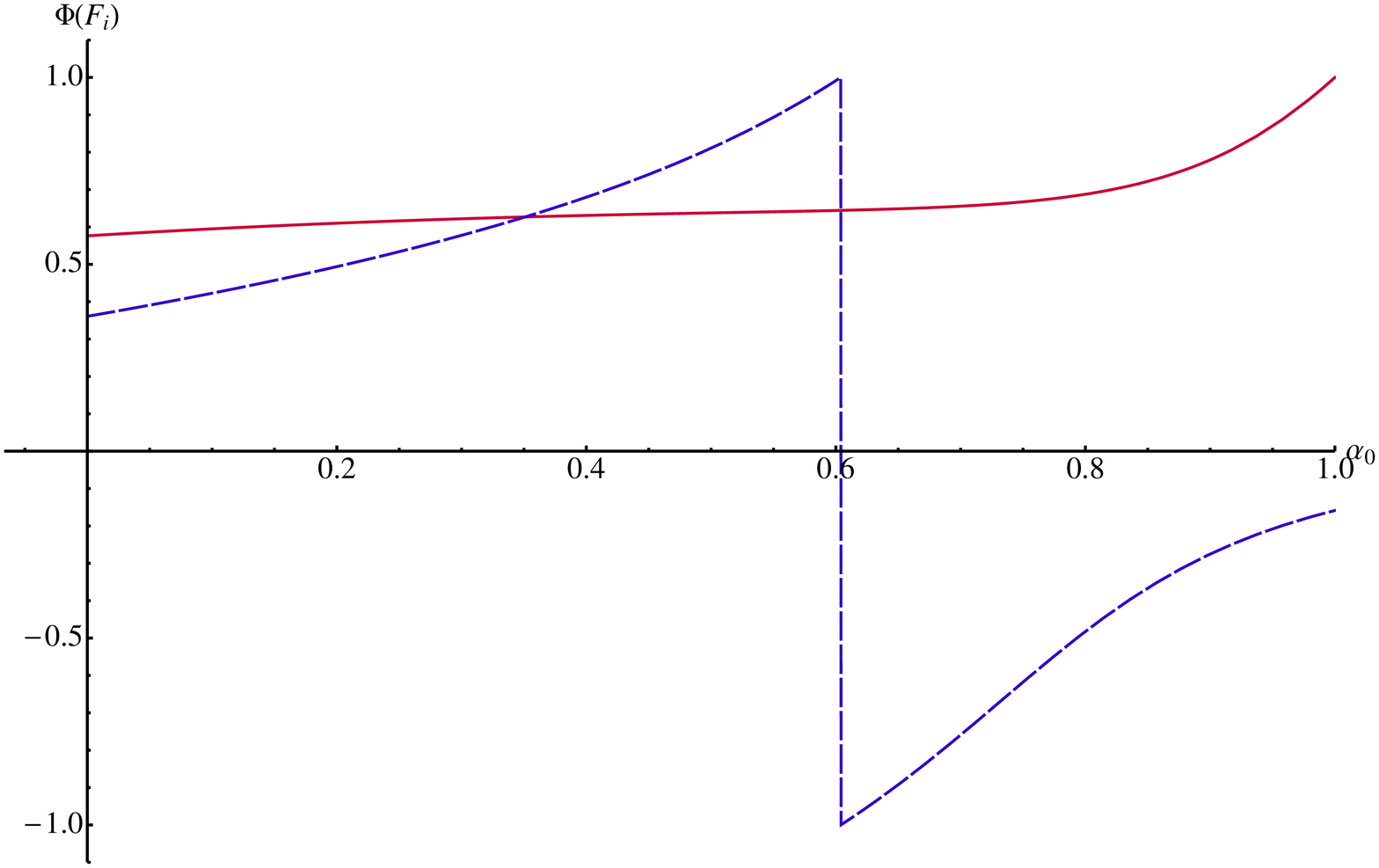} \put(-240,160){(b)\hspace{1cm}\raisebox{-12pt}{$s=2$}} \\[12pt]
    \includegraphics[width=8.5cm]{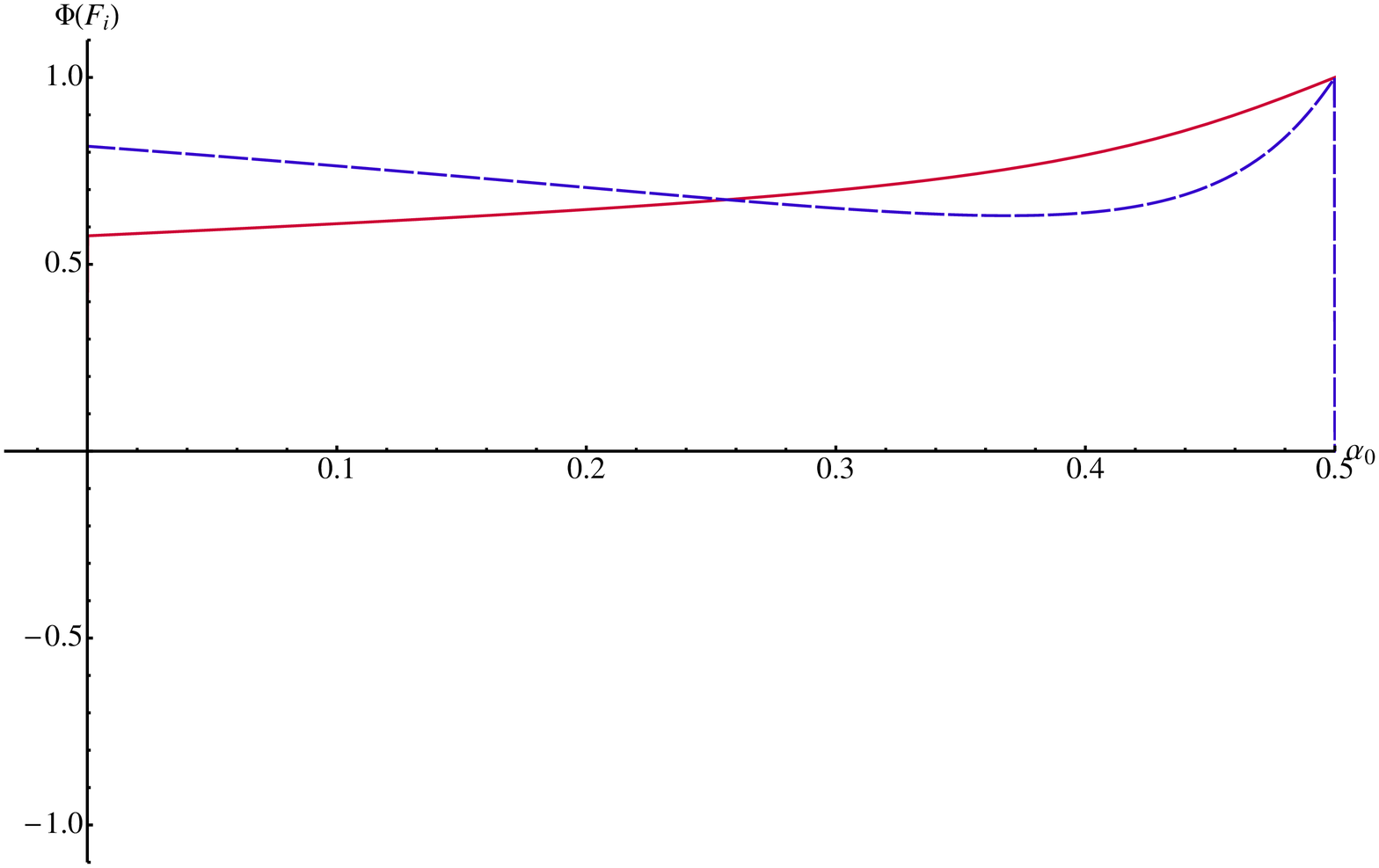} \put(-240,160){(c)\hspace{1cm}\raisebox{-12pt}{$s=1$}}
  \caption{Solution of equations~\eqref{Gamma1}
    and~\eqref{Gamma2} for (a) $s=3$, (b) $s=2$ and (c) $s=1$:
    We display $F_1$ ({\color{red}solid}) and $F_2$
    ({\color{blue}dashed}) for as functions of $\alpha_0$
    with $\alpha_{\rm gh}(\alpha_0)$ from~\eqref{Gamma1}.
    We have employed the mapping
    $\Phi(x)=\frac2{\pi}\arctan\frac{x}{\pi}$ to compactify the
    range $(-\infty,\,\infty)$ to $(-1,1)$. Solutions are determined
    by intersection points. Note that horizontal
    lines correspond to odd-order poles, so there is
    no solution for $s=3$ at $\alpha_0\approx 1.31$
    and no solution for $s=2$ at $\alpha_0\approx 0.605$.}
  \label{fig:IRExp_F1F2}
\end{figure}

We have employed the \texttt{findroot} routine of \emph{Mathematica} 5.2
and 7.0.0 with a wide variety of initial guesses in order to find all branches.
(In the color version different starting points can be recognized for
having different shades of blue and green.) The corresponding
plots are given in Figures~\ref{fig:IRExp_seq3} to~\ref{fig:IRExp_seq1}.
For comparison also the $F_1$ vs. $F_2$ plot for these three cases
is given in Figure~\ref{fig:IRExp_F1F2}.

The two-dimensional plot is particularly interesting in the case $s=2$
(Figure~\ref{fig:IRExp_seq2}), since it shows that the system of equations
is relatively ill-conditioned and the intersection
point (i.e. the solution) would be sensitive even to small perturbations.
(Such perturbations are of course absent in the present truncation,
but could be introduced, for example, by vertex dressing.)

For $s=3$ and $s=2$, the physical values are expected to be
given by the smallest solution for $\alpha_0$, as summarized
in Table~\ref{tab:IRExp_exponents}.

\begin{table}
\begin{tabular}{|c||c|c||c|c|} \hline
 & $\alpha_0$ & $\alpha_{\rm gh}$ & $\alpha_{\bf A}$ & $\alpha_{\pi'}$ \\ \hline
$s=3$ & $1.07945$ & $0.240044$ & $-0.980087$ & $-0.599366$ \\
$s=2$ & $0.351045$ & $0.105460$ & $-1.21092$ & $-0.140125$ \\
$s=1$ & $0.256229$ & $0.068302$ & $-1.63660$ & $-0.119625$ \\ \hline
\end{tabular}
\caption{Infrared exponents, obtained from the smallest solution of
equations~\eqref{Gamma1} and~\eqref{Gamma2} for different values of $s$.
Note that from~\eqref{eq:IRExp_powercountrel}
we have $\alpha_m = \alpha_{\rm gh}$ and $\alpha_\varphi=\alpha_{\pi'}$.}
\label{tab:IRExp_exponents}
\end{table}

For $s=1$ the situation is somehow different since strictly speaking our
equations are not well-defined (because the transverse projectors vanish for $s=1$).
So instead of plainly taking the value obtained for $s=1$ we instead study
the solutions for $s=1+\eps$ with some small, but positive number $\eps$.
For $s=1.06$ there is only one solution at $\alpha_0=0.512478$.

When we lower $s$ to $s\approx 1.0541071910$, in addition to the previous
solution (which has now moved to $\alpha_0\approx 0.51136$) another solution arises at
$\alpha_0\approx 0.17026$. This solution splits into two independent
solutions for smaller values of $s$; the smaller solution approaches
zero for $s\to1^+$, as indicated by the values given in Table~\ref{tab:IRExp_exponentslimit}.

\begin{table}
\begin{tabular}{|c||c|c|c||c|} \hline
 & $\alpha_0^{(1)}$ & $\alpha_0^{(2)}$ & $\alpha_0^{(3)}$ & $\alpha_{\rm gh}^{(1)}$ \\ \hline
$s=1.06$ & $-$ & $-$ & $0.512478$ & $-$ \\
$s=1.05$ & $0.131834$ & $0.202188$ & $0.510570$ & $0.0408666$ \\
$s=1.04$ & $0.093892$ & $0.224380$ & $0.508586$ & $0.0298297$ \\
$s=1.03$ & $0.065958$ & $0.236660$ & $0.506537$ & $0.0212898$\\
$s=1.02$ & $0.041940$ & $0.245122$ & $0.504421$ & $0.0137068$ \\
$s=1.01$ & $0.020197$ & $0.251402$ & $0.502242$ & $0.0066706$ \\ \hline
\end{tabular}
\caption{Infrared exponents, obtained from the smallest
of equations~\eqref{Gamma1} and~\eqref{Gamma2} for
different values of $s$ close to $s=1$. The exponent $\alpha_{\rm gh}$
is only given for the solution close to $\alpha_0=0$. (Note that
relation~\eqref{smallalpha0} is fulfilled quite well for this solution.)}
\label{tab:IRExp_exponentslimit}
\end{table}

\begin{figure}
    \includegraphics[width=8.5cm]{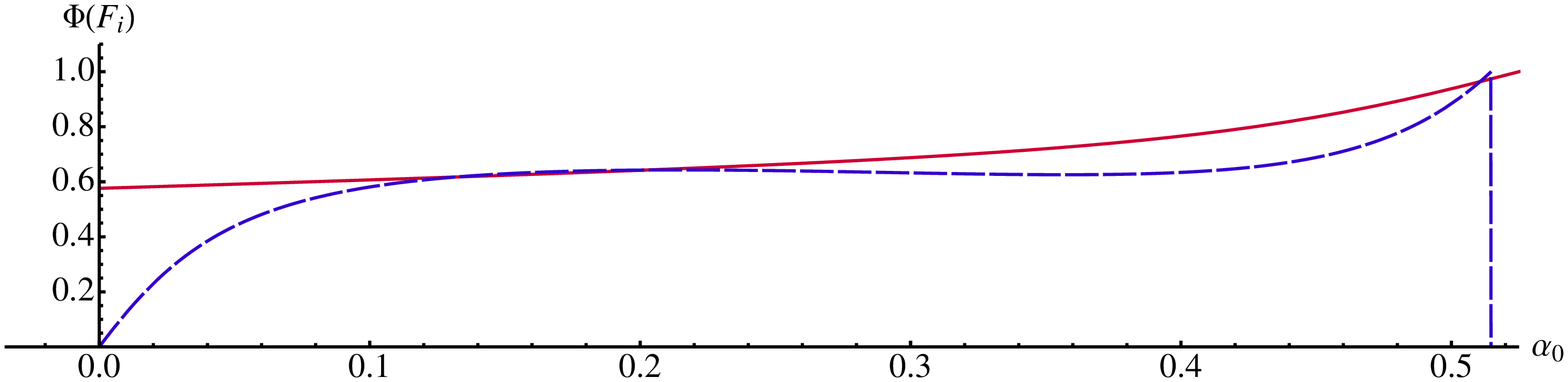}     \put(-240,70){(a)\hspace{1cm}\raisebox{-12pt}{$s=1.05$}} \\[12pt]
    \includegraphics[width=8.5cm]{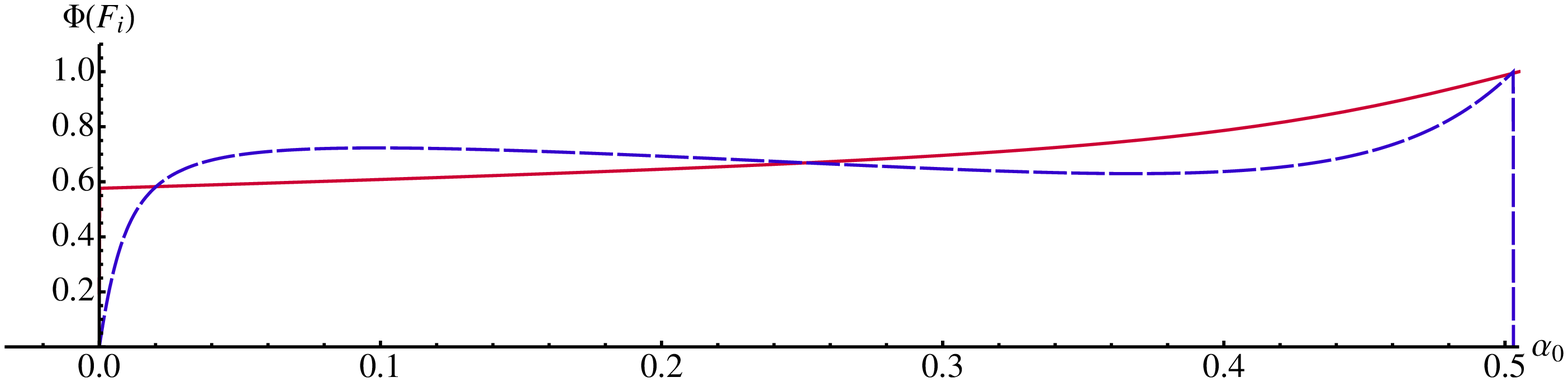}     \put(-240,70){(b)\hspace{1cm}\raisebox{-12pt}{$s=1.01$}} \\[12pt]
    \includegraphics[width=8.5cm]{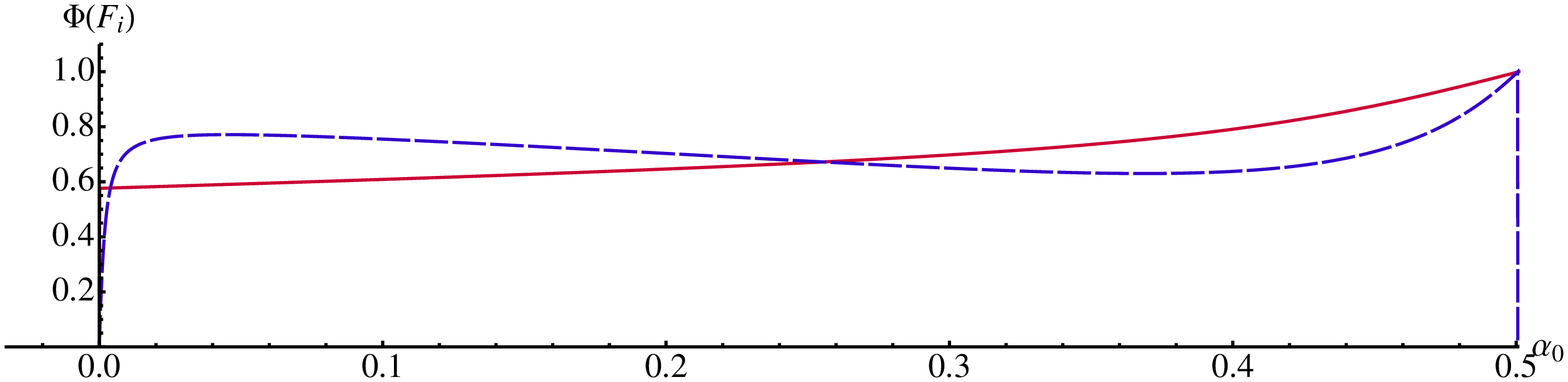}   \put(-240,70){(c)\hspace{1cm}\raisebox{-12pt}{$s=1.002$}} \\[12pt]
    \includegraphics[width=8.5cm]{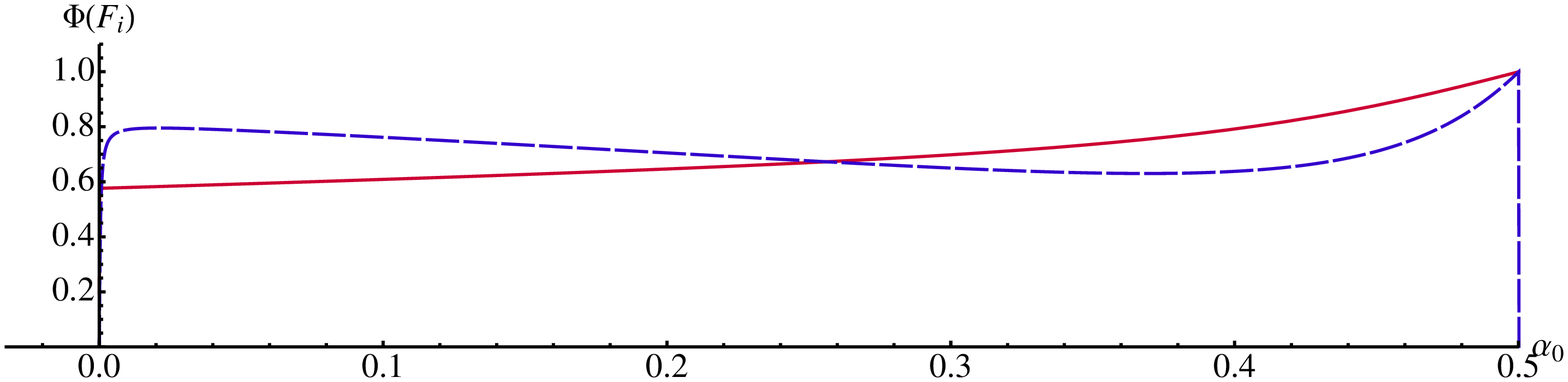} \put(-240,70){(d)\hspace{1cm}\raisebox{-12pt}{$s=1.0004$}}
 \caption{Solution of equations~\eqref{Gamma1}
    and~\eqref{Gamma2} for (a) $s=1.05$, (b) $s=1.01$, (c) $s=1.002$
    and (d) $s=1.0004$, otherwise as in Figure~\ref{fig:IRExp_F1F2}.
    One clearly notes that (as also suggested by Table~\ref{tab:IRExp_exponentslimit})
    there is a solution which approaches $\alpha_0=\alpha_{\rm gh}=0$ for
    $s\to 1^+$ but vanishes for $s=1$.}
  \label{fig:IRExp_F1F2extrapolate}
\end{figure}

This is also illustrated in Figure~\ref{fig:IRExp_F1F2extrapolate} which, together
with Figure~\ref{fig:IRExp_F1F2}.(c) gives a good impression of what is going on:
For $s<1.0541071910$ there is a region around $\alpha_0\approx 0.17026$ where $F_2>F_1$.
For $s=1+\eps$ with $0< \eps<0.0541071910$ we have $F_2(0)=0$ and $F_1(0)>0$,
so there has to be an intersection point. For $s=1$, however, we find $F_2(0)>F_1(0)$.
The family of functions $\{F_2\}_{s=1+\eps,\eps>0}$ seems to converge pointwise,
but not uniformly towards $F_2|_{s=1}$ when $\eps$ approaches zero from above.

\section{Relations among the $b$-coefficients}
\label{sec:IRExpFinT_RelbCoeff}

Having determined the 6 infrared critical exponents, the $\alpha$'s, we turn to the $b$-coefficients.
There are 6 DS equations and 6 coefficients $b_{\bf A}, b_{\pi'}, b_{\rm gh}, b_0, b_m, b_\varphi$.  However,
as we have seen, only 4 of the equations for the $b$'s are independent, so there are 4 relations
satisfied by the $b$'s.  These are, from~\eqref{Gghgh},
\beq
  b_{\bf A} b_{\rm gh}^2 = { - 1 \over I_{S(V, S)} (\alpha_{\bf A}, \alpha_{\rm gh})};
\eeq
from~\eqref{Gpipi}	
\beq
  b_{\bf A} b_0 b_{\pi'}  =  {1 \over I_{V(V, S)} (\alpha_{\bf A}, \alpha_0)};
\eeq
eq.~\eqref{F1}
\beq
  {b_{\pi'} \over b_\varphi } = F_2(\alpha_0);
\eeq
and eq.~\eqref{bghbm},
\begin{equation}
  b_{\rm gh}^2 = b_0 b_\varphi + b_m^2. 
\end{equation}
The last two equations determine the triple products 
\beq
  b_{\bf A} b_0 b_\varphi = {1 \over F_2(\alpha) I_{V(V, S)} (\alpha_{\bf A}, \alpha_0)}
\eeq
\beq
  b_{\bf A} b_m^2 =  {- 1 \over F_2(\alpha) I_{V(V, S)} (\alpha_{\bf A}, \alpha_0)}
    + { - 1 \over I_{S(V, S)} (\alpha_{\bf A}, \alpha_{\rm gh})}.
\eeq

\section{Range of Validity}
\label{sec:IRExpFinT_RangVal}

So far we have only taken into account the zeroth Matsubara frequency,
thus working effectively in the infinite-temperature limit. However, since
infrared properties of the theory are governed by the zero frequency
contribution also at finite temperatures (for all $T\ne 0$, even those in
the confined phase) the results obtained so far may more general than
initially stated.

All Matsubara frequencies $\omega_n$ with $n\ge 1$ effectively
introduce mass terms in the non-instantaneous propagators, so
all such contributions decouple from the deep infrared where
critical exponents are valid. Also in loop terms massive contributions
show up only pairwise, so taking into account only the $0^{\rm th}$
Matsubara frequency yields a closed system.

The only instance where zero and nonzero Matsubara frequencies
could directly be intertwined is imposing the horizon condition --
an issue which should be the subject of closer investigation.


\section{Discussion and Summary}
\label{sec:IRExpFinT_Summary}

\subsection{Discussion of the Infrared Exponents}
\label{sec:IRExpFinT_DiscIRExp}

We now turn to the discussion of the results given
in table~\ref{tab:IRExp_exponents}. The horizon condition
tells us that the ghost propagator
$D_{\rm gh} \sim {b_{\rm gh} \over (k^2)^{1 + \alpha_{\rm gh} } }$
is enhanced in the infrared, so we are interested in
solutions which fulfill
\[
  \alpha_{\rm gh} > 0\,.
\]
As one sees from the black curves in Figs. 2, 3 and 4, this is true for our solution in the whole interval $0 < \alpha_0 <\frac s 2$
which is the interval of physical interest.

The most interesting quantity in Coulomb gauge
Yang-Mills theory is presumably the color-Coulomb potential, given
in momentum space by
\beq
  D_{A_0A_0}(k) \sim {{b_0m^{2\alpha_0}} \over (k^2)^{1 + \alpha_0} }.
\eeq
It is linearly rising in position space for $\alpha_0|_s = (s - 1)/2$.

For $s = 3$, this gives $\alpha_0|_3 = 1$, and the result we obtain lies
slightly above this value, at $\alpha_0 = 1.07945$. This corresponds
to a slightly more than linearly rising potential\footnote{Note that
the color-Coulomb potential is not a gauge-invariant quantity,
so the arguments which forbid more-than linearly rising
potentials in relativistic field theory do not directly apply;
also the asymptotic inequality between Wilson and
color-Coulomb potential in~\cite{Zwanziger:2003}
would be satisfied. As discussed in section~\ref{sec:IRExpFinT_IRDiv},
the mathematical aspects of such potentials are under
control. Still it would be very surprising if in reality
$D_{A_0A_0}$ were more than linearly rising.}.  The transverse gluon propagator $D_{\bf AA}(k)$ vanishes at
$k = 0$ if $\alpha_{\bf A} < -1$. From Table~\ref{tab:IRExp_exponents} we have $\alpha_{\bf A}|_{s=3}=-0.980087$, which corresponds to $D_{\bf AA}(k)$ that is weakly divergent at $k = 0$.  This value  is close to $\alpha_{\bf A}=-1$, which corresponds to a finite value for $D_{\bf AA}(k = 0)$, and a small change caused by an improved truncation could also change this to $\alpha_{\bf A} < -1$, corresponding to  a gluon propagator $D_{\bf AA}(k)$ that vanishes at $k = 0$.

For $s=2$, the value from Table~\ref{tab:IRExp_exponents}, $\alpha_0 = 0.351045$, lies below the linearly rising case, $\alpha_0 = (s - 1)/2 = 0.5$.  However as can be seen from Figure~\ref{fig:IRExp_seq2}, the system
is badly conditioned so that, if properly dressed vertices or terms neglected in our truncation even slightly modify the system of equations, then any value $\alpha_0\in(0,\,0.55)$ [which determines a value of $\alpha_{\rm gh} \in (0, \, 0.146)$] could qualify as a possible solution. This includes the
linearly rising case. From Table~\ref{tab:IRExp_exponents} we find $\alpha_{\rm A}=-1.21092<-1$,
which implies that $D_{\bf AA}(k)$ vanishes at $k=0$, but the uncertainty of
$\alpha_0$, due to ill-conditioning, extends to all critical exponents, including $\alpha_{\rm A}$. 

For $s = 1$, the system is exactly solvable analytically.  Since our equations do not strictly apply at $s = 1$ (see discussion in Sec.~\ref{ssec:IRExp_numres}) we have examined numerically the limit $s \to 1^+$ (see Table~\ref{tab:IRExp_exponentslimit} and Figure~\ref{fig:IRExp_F1F2extrapolate}). The results obtained this way are consistent with eq.~\eqref{smallalpha0} which is valid for small $\alpha_0$.  For $s\to1^+$ this solution converges towards
\beq
  \alpha_0 = \alpha_{\rm gh} = 0.
\eeq
These values agree with the analytic solution presented in Appendix B.  By~\eqref{eq:IRExp_powercountrel} this would correspond to $\alpha_A=-\frac32$,
but transverse propagators are not well-defined for $s=1$, so this infrared critical exponent is undefined for the theory at $s = 1$. 

	Because of the symmetry~\eqref{symofbs} the coefficient $b_0$ is undetermined in the infrared asymptotic limit and must be fixed by subdominant terms that we have neglected.  This is unfortunate because according to~\eqref{eq:massscale} and~\eqref{eq:powerlawansatz}, for $\alpha_0=1$, the quantity $b_0(g^2T)^{2\alpha_0}$ represents the color-Coulomb string tension and one would have liked to make a comparison with lattice determinations of this quantity~\cite{Greensite:2004ke, Nakagawa:2006fk,Nakagawa:2007zzc, Nakagawa:2007zzb,Voigt:2008rr}.

\subsection{Conclusion}

An ideal theory of confinement would explain in simple terms why there is a linearly rising Wilson potential and, in the Coulomb-gauge scenario, why there is a linearly rising color-Coulomb potential, with a Coulomb string tension $\sigma_{coul}(T)$ that increases with $T$ even in the deconfined phase~\cite{Greensite:2004ke}.  This we have not achieved; our solution of the Coulomb-gauge DSE at high temperature is obtained from equations involving $\Gamma$-functions that must be solved numerically.  Nevertheless it remains true that the numerical value obtained for $s = 3$ space dimensions is numerically close to a linearly rising potential, with $V(r) \sim r^{2 + 2 \alpha_0 -s} = r^{1.15890}$. 

In $s =2$ spatial dimensions the agreement is not as good but this may be due to the ill-conditioning of the equations. Finally, as $s$ approaches 1 spatial dimension, the smallest solution approaches the exact analytic result.

\medskip


\bigskip

{\bf Acknowledgements}\\

\noindent K.L. was supported by the Doktoratskolleg \emph{Hadronen
im Vakuum, in Kernen und Sternen} (FWF DK W1203-N08)  and by the
\emph{Graz Advanced School of Science} (NAWI-GASS). He would like
to  express his thanks for the hospitality of New York University (NYU),
where a considerable part of this work has been done.

The authors are grateful to Reinhard Alkofer and Axel Maas for
valuable comments and discussions.

\appendix

\section{Evaluation of Power-Law Integrals}
\label{app:IRExpFinT_PowerLaw}

Determination of the infrared critical exponents requires evaluation
of the integrals~\eqref{ISVS} to~\eqref{IVSS} which yields the results
stated in~\eqref{valISVS} to~\eqref{valIVSS}. In order to illustrate the
calculational scheme, we explicitly discuss the evaluation the
integral~\eqref{ISVS}. [The evaluation of such integrals is also
discussed for example in the appendix of~\cite{Zwanziger:2001kw};
here we give some more details on intermediate steps.]

First we check issues of convergence: We write
\begin{align}
  I_{S(V,S)}(\alpha_{\bf A}, \alpha_\varphi) &=
  N k^{-s +2 \alpha_{\bf A} + 2 \alpha_\varphi + 2}\,I_1
\end{align}
with
\beq
  I_1 := \int { d^s p \over (2 \pi)^s } \ 
  {k^2 p^2 - (p \cdot k)^2 \over (p^2)^{2+\alpha_{\bf A}} \ [(k - p)^{2}]^{1 + \alpha_\varphi}}\,.
  \label{eq:app_ISVSint}
\eeq
In this integral we obtain the power $p^{s+2}$ from the numerator.
In the infrared ($p\to 0$) the term $(k - p)^2$ is finite for $k\ne 0$,
so we can neglect it for questions of convergence.\footnote{Note
that for $k=0$ the numerator vanishes, so this case is not problematic;
for all other cases the argument holds.} Thus the denomiator contributes
the power $p^{2(2+\alpha_{\bf A})}$ and a necessary condition for infrared
convergence is
\beq
  s+2>4 + 2\alpha_{\bf A}\,,\qquad\text{i.e.}\qquad \alpha_{\bf A}<\frac{s}2-1\,.
\eeq
\eqref{eq:app_ISVSint} may also have a non-integrable singularity at $p=k$.
By the substitution $p\to p+k$ the integral takes the form
\beq
  I_1=\int { d^s p \over (2 \pi)^s } \ 
  {k^2 p^2 - (p \cdot k)^2 \over [(p+k)^2]^{2+\alpha_{\bf A}} \ (p^{2})^{1 + \alpha_\varphi}}\,.
\eeq
The denomiator now contributes the power $p^{2(1+ \alpha_\varphi)}$
and we see that we also have to demand
\beq
  s+2>2 + 2\alpha_\varphi\,,\qquad\text{i.e.}\qquad \alpha_\varphi<\frac{s}2
\eeq
in order to have infrared convergence. In the ultraviolet, $k$~is negligible compared
to~$p$ and the denomiator in~\eqref{eq:app_ISVSint} contributes the power
$p^{2(2+\alpha_{\bf A}+2(1+ \alpha_\varphi))}$. To have ultraviolet convergence
we have to demand
\beq
  s+2<6+2\alpha_{\bf A}+2\alpha_\varphi\,,
  \quad\text{i.e.}\quad
  \alpha_{\bf A}+\alpha_\varphi > \frac{s}2-1.
\eeq
These conditions are compatible; by combining them one can also deduce
the conditions $\alpha_\varphi>0$ and $\alpha_{\bf A}>-1$. 

\smallskip

Now we proceed with the evaluation of the integrals.
In~\eqref{eq:app_ISVSint} we represent the powers of
propagators by integrals over auxiliary variables.
This employs the integral representation of the gamma function,
\beq
   \int_0^{\infty} t^{x-1}\,\E^{-kt}\,\d t \;\overset{u=kt}=\;\frac1{k^x}\int_0^{\infty} u^{x-1}\,\E^{-u}\,\d u
     = \frac{\Gamma(x)}{k^x}\,,
\eeq
($k>0$) which allows us to write
\begin{align}
  \frac1{(p^2)^{2+\alpha_{\bf A}}} &=
    \int\nolimits_0^{\infty} \d a\, \frac{a^{1+\alpha_{\bf A}}}{\Gamma(2+\alpha_{\bf A})}\,\E^{-a\,p^2}\,, \\
  \frac1{[(k - p)^{2}]^{1 + \alpha_\varphi}}&=
    \int\nolimits_0^{\infty} \d b\, \frac{b^{\alpha_\varphi}}{\Gamma(1+\alpha_\varphi)}\,\E^{-b\,(k-p)^2}\,.
\end{align}
With these identities $I_1$ takes the form
\beq
  I_1 = \int\nolimits_0^{\infty} \d a\, \frac{a^{1+\alpha_{\bf A}}}{\Gamma(2+\alpha_{\bf A})}
    \int\nolimits_0^{\infty} \d b\, \frac{b^{\alpha_\varphi}}{\Gamma(1+\alpha_\varphi)} I_2 \\
\eeq
with
\begin{align}
  I_2 &= \int { d^s p \over (2 \pi)^s }\,\left[ k^2 p^2 - (p \cdot k)^2 \right]\,\E^{-\Phi}\,,\\
  \Phi &= ap^2 + b(k-p)^2 = (a+b)p^2 + bk^2 - 2b(k\cdot p) \nonumber \\
  &=(a+b)\left(p-{\textstyle\frac{b}{a+b}}\,k\right)^2 - {\textstyle\frac{b^2}{a+b}}\,k^2 + bk^2 \nonumber \\
  &=:(a+b)q^2+{\textstyle\frac{ab}{a+b}}\,k^2\,.
\end{align}
The substitution $p\to q=p-{\textstyle\frac{b}{a+b}}\,k$ yields
\begin{align}
  I_2 &=  \int { d^s q \over (2 \pi)^s }\, (k^2 q^2 - (q \cdot k)^2) \,\E^{-(a+b)q^2-\frac{ab}{a+b}k^2}\,.
     \nonumber \\
  &=\E^{-\frac{ab}{a+b}k^2}
     \,k^2\! \int {\d^s q \over (2 \pi)^s }\ (1 - (\hat{q} \cdot \hat{k})^2) \,q^2 \,\E^{-(a+b)q^2}\,.
\end{align}
One can rewrite an integral of the form
\beq
  J=k^2\! \int {\d^s q \over (2 \pi)^s }\, (1 - (\hat{q} \cdot \hat{k})^2) \,f(q^2)
\eeq
as
\beq
  J=k_ik_j\! \int {\d^s q \over (2 \pi)^s }\, (\delta_{ij} - \hat{q}_i\hat{q}_j) \,f(q^2)\,.
\eeq
Since $J$ is a scalar which is quadratic in $k$, one has to find
\beq
  \int {\d^s q \over (2 \pi)^s }\, (\delta_{ij} - \hat{q}_i\hat{q}_j) \,f(q^2) = C\,\delta_{ij}\,.
\eeq
Taking the trace of this equation ($\delta_{ii}=s$, $\hat{q}_i\hat{q}_i=1$)
and a small rearrangement of factors yields
\beq
  C = \frac{s-1}{s}\int {\d^s q \over (2 \pi)^s }\,f(q^2)\,.
\eeq
So we have 
\beq
  I_2 =\E^{-\frac{ab}{a+b}k^2}\,k^2\,\frac{s-1}{s}\,I_3
\eeq
with
\begin{align}
  I_3 &=\int {\d^s q \over (2 \pi)^s }\, q^2 \,\E^{-(a+b)q^2} \nonumber \\
  &=\frac1{(2 \pi)^s} \int\d\Omega_s\int\nolimits_0^{\infty}\,q^{s+1} \,\E^{-(a+b)q^2}
\end{align}
The substitution $x=(a+b)q^2$ yields
\begin{align}
  I_3 &=\frac1{(2 \pi)^s} \int\d\Omega_s
    \int\nolimits_0^{\infty} \left( \frac{x}{a+b} \right)^{(s+1)/2}\frac{\E^{-x}\,\d x}{2x^{1/2}(a+b)^{1/2}} \nonumber \\
 &=\frac1{(2 \pi)^s}\int\d\Omega_s\,\frac1{2(a+b)^{1+s/2}}
    \int\nolimits_0^{\infty} x^{(s/2+1)-1}\,\E^{-x}\,\d x \nonumber \\
 &= \frac1{(2 \pi)^s}\,\frac{2\,\pi^{s/2}}{\Gamma(\frac{s}2)}\,\frac1{2(a+b)^{1+s/2}}
    \Gamma\left(\frac{s}2+1\right) \nonumber \\
 &= \frac1{(4\pi)^{s/2}}\,\frac1{(a+b)^{1+s/2}}
   \frac{ \frac{s}2\,\Gamma\left(\frac{s}2\right)}{\Gamma(\frac{s}2)} \nonumber \\
 &= \frac{s}{2\,(4\pi)^{s/2}\,(a+b)^{1+s/2}}
\end{align}
Consequently we obtain (with slight rearrangements)
\begin{align}
  I_2 &=\frac{(s-1)\,k^2}{2(4\pi)^{s/2}}\,
  \frac{\E^{-\frac{ab}{a+b}k^2}}{(a+b)^{s/2+1}}\,.
\end{align}
To proceed with the evaluation of
\begin{align}
  I_1 &= \frac{(s-1)\,k^2}{2(4\pi)^{s/2}}\,
  \int\nolimits_0^{\infty} \d a\, \frac{a^{1+\alpha_{\bf A}}}{\Gamma(2+\alpha_{\bf A})} \nonumber \\
   &\quad\times \int\nolimits_0^{\infty} \d b\, \frac{b^{\alpha_\varphi}}{\Gamma(1+\alpha_\varphi)}\,
    \frac{\E^{-\frac{ab}{a+b}k^2}}{(a+b)^{s/2+1}}\,.
\end{align}
we introduce another auxiliary variable $\tau$ by integrating over $\delta(a+b-\tau)$,
\begin{align}
  I_1 &= \frac{(s-1)\,k^2}{2(4\pi)^{s/2}}\,\int\nolimits_0^{\infty} \d \tau
  \int\nolimits_0^{\infty} \d a\, \frac{a^{1+\alpha_{\bf A}}}{\Gamma(2+\alpha_{\bf A})} \nonumber \\
   &\quad\times \int\nolimits_0^{\infty} \d b\, \frac{b^{\alpha_\varphi}}{\Gamma(1+\alpha_\varphi)}\,
    \frac{\E^{-\frac{ab}{a+b}k^2}}{(a+b)^{s/2+1}}\,\delta(a+b-\tau)\,.
\end{align}
The substitution $a=\tau\alpha$, $b=\tau\beta$ yields
\begin{align}
  I_1 &= \frac{(s-1)\,k^2}{2(4\pi)^{s/2}}\,\frac1{\Gamma(2+\alpha_{\bf A})\,\Gamma(1+\alpha_\varphi)}
  \int\nolimits_0^{\infty} \d\alpha \, \int\nolimits_0^{\infty} \d\beta \nonumber \\
   &\quad\times \frac{\delta(\alpha+\beta-1)}{(\alpha+\beta)^{s/2+1}}
   \,\alpha^{1+\alpha_{\bf A}}\beta^{\alpha_\varphi}\,I_4
\end{align}
with
\begin{align}
   I_4 &:= \int\nolimits_0^{\infty} \d \tau \,\tau^{\alpha_{\bf A}+\alpha_\varphi+1-\frac{s}2}\,\E^{-\alpha\beta\tau\,k^2}\,.
\end{align}
(Note that $\delta(\tau\alpha+\tau\beta-\tau)=\frac1{\tau}\delta(\alpha+\beta-1)$.)
We now substitute $t=\alpha\beta\,k^2\,\tau$,
\begin{align}
   I_4 &:= \frac1{(\alpha\beta\,k^2)^{\alpha_{\bf A}+\alpha_\varphi+2-\frac{s}2}}
   \int\nolimits_0^{\infty} \d t \,t^{(\alpha_{\bf A}+\alpha_\varphi+2-\frac{s}2)-1}\,\E^{-t} \nonumber \\
   &=\frac{\Gamma(\alpha_{\bf A}+\alpha_\varphi+2-\frac{s}2)}%
   {(\alpha\beta)^{\alpha_{\bf A}+\alpha_\varphi+2-\frac{s}2}\;k^{2\alpha_{\bf A}+2\alpha_\varphi+4-s}}\,.
\end{align}
and obtain
\beq
  I_{S(V,S)}(\alpha_{\bf A}, \alpha_\varphi) =
  \frac{N(s-1)}{2(4\pi)^{s/2}}\,\frac{\Gamma(\alpha_{\bf A}+\alpha_\varphi+2-\frac{s}2)}%
  {\Gamma(2+\alpha_{\bf A})\,\Gamma(1+\alpha_\varphi)}\,I_5
  \label{eq:app_ISVSwoBeta}
\eeq
with
\beq
  I_5= \int\nolimits_0^{\infty} \d\alpha \, \int\nolimits_0^{\infty} \d\beta \,
 \frac{\delta(\alpha+\beta-1)}{(\alpha+\beta)^{s/2+1}} \,
 \alpha^{\frac{s}2 - 1 - \alpha_\varphi}
 \beta^{\frac{s}2 -2 -\alpha_{\bf A}}.
 \label{eq:app_almostbeta}
\eeq
Because of the delta functional, one only has contributions from $\alpha+\beta\le1$,
accordingly one can restrict the integrations in~\eqref{eq:app_almostbeta} to the
interval $[0,\,1]$. Performing one integration explicitly yields
\begin{align}
    I_5
   &=\int\nolimits_0^1 \d\alpha \, \alpha^{(\frac{s}2 - \alpha_\varphi) - 1}
   (1-\alpha)^{(\frac{s}2 -1 -\alpha_{\bf A})-1} \nonumber \\
   &=\mathrm{B}\left(\frac{s}2 - \alpha_\varphi,\,\frac{s}2 -1 -\alpha_{\bf A}\right) \nonumber \\
   &=\frac{\Gamma(\frac{s}2 - \alpha_\varphi)\,\Gamma(\frac{s}2 -1 -\alpha_{\bf A})}%
   {\Gamma(s-1- \alpha_\varphi-\alpha_{\bf A})}\,.
\end{align}
Plugging this result into~\eqref{eq:app_ISVSwoBeta} yields
\begin{align}
I_{S(V,S)}(\alpha_{\bf A}, \alpha_{\rm gh}) &= 
{N (s-1) \over 2 (4 \pi)^{s/2}} \  
{ \Gamma(\alpha_{\bf A}+\alpha_\varphi+2-\frac{s}2)
 \over 
 \Gamma(2 + \alpha_{\bf A})  }
\nonumber  \\
&\quad\times {  \Gamma(\frac{s}2 - \alpha_\varphi)\,\Gamma(\frac{s}2 -1 -\alpha_{\bf A})
 \over 
 \Gamma(1 + \alpha_{\rm gh})  \ \Gamma(s - 1 - \alpha_{\rm gh} - \alpha_{\bf A}) }\,,
\end{align}
which is the result stated in~\eqref{valISVS}. The other power-law integrals
are evaluated in a similar way.
Note that the integrals $I_{V(\cdot,\cdot)}$ contain a factor $(s-1)^{-1}$
which stems from taking the trace of the projector on the lhs of the
corresponding DS equation,
\beq
  \mathrm{tr}P_{ij}(p) = \mathrm{tr}\left(\delta_{ij}-\frac{p_ip_j}{\V{p}^2}\right)
  = \delta_{ii}-\frac{p_ip_i}{\V{p}^2} =  s - 1\,,
\eeq
and is included for convenience in the definition of the integral.

\section{Analytic Result for $s=1$}
\label{app:IRExpFinT_Analytic}

	We start from the second-order action (\ref{secondorder}) in $s = 1$ space dimensions,
\beq
S_1 = {1 \over T} \int \d x_1 \ \left( {1 \over 2} (D_1A_0)^2 
    + \p_1\bar{c}D_1 c\right).
\eeq 
where $D_1(A) = \p_1 + gA_1 \times$, and the Coulomb gauge condition reads $\p_1 A_1 = 0$.  Thus $A_1$ is independent of $x_1$.
The propagators of $A_0$ and the ghost are given by
\beqa
\langle A_0(x_1) A_0(y_1) \rangle & = & \Big\langle {T \over - D_1^2} \Big\rangle 
\nonumber \\
\langle c(x_1) \bar c (y_1) \rangle & = & \Big\langle {T \over - \p_1 D_1} \Big\rangle 
\eeqa
where the average is with respect to $A_1$.  Since $A_1$ is independent of $x_1$, we may diagonalize these operators by Fourier transform, so
\beqa
D_{A_0 A_0}(k_1) & = & \Big\langle {T \over (k_1 + \I g A_1\times)^2} \Big\rangle 
\nonumber \\
D_{\rm gh}(k_1) & = & \Big\langle {T \over k_1 (k_1 + \I g A_1\times)} \Big\rangle 
\eeqa

We now integrate $A_1$ over the Gribov region, which is the region where the eigenvalues of the Faddeev-Popov operator $k_1(k_1 + igA_1\times)$ are non-negative.  For this purpose, we quantize in a periodic box of length $L$, so $k_1 = 2\pi n/L$, where $n$ is an integer.  We first consider the gauge group to be the SU(2) group.  One easily finds that the eigenvalues of $\I gA_1 \times $ are given by $0$ and $\pm |g A_1|$.  The case $k_1 = 0$ is trivial.  The Gribov horizon is determined by the first non-trivial zero eigenvalue.  This occurs for $n = \pm 1$, at $ |g A| = 2 \pi / L$.  Thus $A_1$ is integrated over the sphere $|A_1| \leq 2 \pi / g L$.  We now take the infinite-volume limit $L \to \infty$, while keeping a typical momentum $k_1$ finite.  In this case we have $A_1 \to 0$, and the $A_0$ and ghost propagators approach their free values, $D_{A_0 A_0}(k_1) = D_{\rm gh}(k_1)  = T/k_1^2$.  The case of SU($N$) is the same.  Accordingly the infrared critical exponents are given by $\alpha_0 = \alpha_{\rm gh} = 0$.  The restriction to the Gribov region (which in this case coincides with the Fundamental Modular Region, where gauge-fixing is unique) was essential in deriving this result.


\end{document}